\documentclass{jfm}
\usepackage{graphicx}
\usepackage{epstopdf, epsfig}
\usepackage{booktabs,multirow}
\usepackage{hyperref}
\def\citeapos#1{\citeauthor{#1}'s (\citeyear{#1})}

\hypersetup{colorlinks,linkcolor={blue}, citecolor={blue}}

\shorttitle{Scalings of scale-by-scale turbulence energy}
\shortauthor{J. G. Chen and J. C. Vassilicos}

\title{Scalings of scale-by-scale turbulence energy in non-homogeneous
turbulence}

\author{J. G. Chen\aff{1}
 \and J. C. Vassilicos\aff{1} \corresp{john-christos.vassilicos@centralelille.fr}}

\affiliation{\aff{1}Univ. Lille, CNRS, ONERA, Arts et Metiers
  Institute of Technology, Centrale Lille, \\
  UMR 9014 - LMFL - Laboratoire de M\'{e}canique des Fluides de Lille - Kamp\'{e} de F\'{e}riet,
  F-59000 Lille, France}

\begin{document}

\maketitle

\begin{abstract}
A theory of non-homogeneous turbulence is developed and is applied to
boundary-free shear flows. The theory introduces assumptions of inner
and outer similarity for the non-homogeneity of two-point statistics
and predicts power law scalings of second order structure functions
which have some similarities with but also some differences from
Kolmogorov scalings. These scalings arise as a consequence of these
assumptions, of the general inter-scale and inter-space energy balance
and of an inner-outer equivalence hypothesis for turbulence
dissipation. They reduce to usual Kolmogorov scalings in stationary
homogeneous turbulence. Comparisons with structure function data from
three qualitatively different turbulent wakes provide support for the
theory’s predictions but also raise new questions for future research.
\end{abstract}


\section{Introduction}

Eighty years ago, \citet{Kolmogorov1941dissipation,
  Kolmogorov1941local} and \citet{obukhov1941distribution} introduced
the theory of homogeneous equilibrium turbulence which became very
influential after \citeapos{batchelor1953theory} monograph on this
theory was published twelve years later. A further fifteen to twenty
years down the line, this theory became a theoretical pillar of
one-point two-equation models (such as the very widely used
$k-\varepsilon$ model) and of two-point subgrid turbulence modelling
for Large Eddy Simulations \citep[e.g. see][]{pope2000}. These
one-point and two-point modelling approaches are used for a wide range
of turbulent flows, well beyond homogeneous equilibrium turbulence
which is in fact of very limited practical interest itself given that
turbulent flows in nature and engineering are typically
non-homogeneous and out of Kolmogorov equilibrium.

The Kolmogorov-Obukhov theory predicts that, in the inertial range,
the turbulence energy spectrum is proportional to the $-5/3$ power of
wavenumber \citep{obukhov1941distribution} and that the second order
structure function is proportional to the $2/3$ power of two-point
separation distance \citep{Kolmogorov1941local}. Quoting
\cite{kraichnan1974kolmogorov}, the $2/3$ and $-5/3$ power laws of
Kolmogorov and Obukhov have ``achieved an embarassment of success'' as
they have ``been found not only where they reasonably could be
expected'' but also where the Kolmogorov theory is not designed for,
e.g. in significantly non-homogeneous and near-field
turbulence. \citet{alves_portela2017} listed some examples of papers
dating from the 1960s onwards which reported observations of $-5/3$
spectra in conditions where the Kolmogorov theory's premises and/or
assumptions do not hold. They also reported that the best-defined
$-5/3$ spectra in the turbulent wake of a square prism are found in
the most inhomogeneous region of the flow's near field, specifically
at a distance equal to about two square prism sizes from the
wake-generating prism. \citet{gomes2014evolution} observed with
Particle Image Velocimetry (PIV) a very well-defined $2/3$ second
order structure function (see their figure 16) at the very near field
of a turbulence-generating grid where the turbulence is not decaying
yet with streamwise distance.

These observations suggest that the Kolmogorov theory of homogeneous
turbulence equilibrium cascade and of the resulting scale-by-scale
energy must be a special case of a more general turbulence theory. To
break beyond the Kolmogorov theory in a substantial way we need to
break beyond its most constraining premises: equilibrium and
homogeneity. Some progress has been made on non-equilibrium cascades
and non-equilibrium dissipation and their consequences on basic flow
profiles over the past ten years, a most recent development having
been contributed by \citet{ortiz2021high}. The present paper attempts
to break beyond the Kolmogorov theory by taking substantial account of
non-homogeneity and of the various energy transport processes that
non-homogeneity involves.

\cite{chen2021turbulence} found a turbulence dissipation
non-homogeneity scaling which is the same in three different turbulent
wake flows. The mechanism of turbulence dissipation in free shear
flows being the non-linear turbulence interscale energy transfer,
their finding would suggest some potential structure of some
generality in cascades which occur in non-homogeneous
turbulence. Their work is therefore an important motivation for the
present paper.

\cite{chen2021turbulence} used two Particle Image Velocimetry (PIV)
set-ups, one with small fields of view for turbulence dissipation
measurements and one with a large field of view for integral length
scale measurements, to study cross-stream profiles of turbulence
dissipation $\varepsilon$, integral length scale $L$ and turbulent
kinetic energy $k$ in the wakes of three different side-by-side pairs
of identical square prisms. Depending on gap-ratio $G/H$, where $G$ is
distance between the two prisms and $H$ is the size of the prisms, the
flow structures and dynamics of such flows can be very different. The
three different values of $G/H$ in the experiments of
\cite{chen2021turbulence} produced three different flow regimes with
significant qualitative differences in dynamics, large-scale features
and turbulent flow non-homogeneity. Yet, in every location that they
sampled in each one of these three flows, they found that the local
turbulence dissipation coefficient $C_{\varepsilon} \equiv \varepsilon
L/k^{3/2}$ and the local Reynolds number $Re_{\lambda} \equiv
\sqrt{k}\lambda/\nu$ (where $\lambda$ is the Taylor length-scale and
$\nu$ is the fluid's kinematic viscosity) are not constant along the
cross-stream coordinate $y$ and vary in opposition to each other:
where one increases with $y$ the other decreases with $y$ and {\it
  vice versa}. After removal of the large-scale coherent structures by
a snapshot Proper Orthogonal Decomposition, the turbulence dissipation
coefficient $C^\prime_{\varepsilon}$ of the remaining incoherent
turbulence fluctuations was found to be independent of viscosity and
proportional to ${Re^\prime}_{\lambda}^{-3/2}$ (where
${Re^\prime}_{\lambda}$ is the Taylor length-based Reynolds number of
the incoherent turbulence) at all streamwise positions tested, in all
three flows for three different inlet Reynolds numbers.

Such universal behaviour is reminiscent of the non-equilibrium
dissipation scaling $C_{\varepsilon}\sim
(\sqrt{Re_{G}}/Re_{\lambda})^{n}$ which appears in non-stationary
conditions either directly in time as in decaying and more generally
time-evolving periodic turbulence (where $n=1$ \citep{goto2015energy,goto2016local,goto2016unsteady}) or in the streamwise direction of decaying
grid-generated turbulence (where $n=1$ \citep{Vassilicos2015}), various
bluff body wakes (where $n=1$ \citep{Vassilicos2015,dairay2015non,obligado2016nonequilibrium,chongsiripinyo_sarkar_2020}), turbulent jets
(where $n=1$ \citep{cafiero2019non}) and slender body wakes
(where $n=4/3$ \citep{ortiz2021high}). In all these
cases, $C_{\varepsilon}$ and $Re_{\lambda}$ vary in time or along the
streamwise coordinate $x$ in opposition to each other: when one
increases the other decreases with $x$ and {\it vice versa}. This is similar
to the observation of \citet{chen2021turbulence}, except that their observation
was made in the transverse/cross-stream direction and is therefore
characteristic of turbulence non-homogeneity rather than
non-stationarity. Also, the value of the exponent $n$ found by \cite{chen2021turbulence} after removal of the large-scale coherent structures is
$n=3/2$.

The non-constancy of $C_{\varepsilon}$ in non-stationary conditions
and the fact that $C_{\varepsilon}$ increases/decreases when
$Re_{\lambda}$ decreases/increases was shown by \cite{goto2016local}
to be the consequence of a non-equilibrium cascade. Could it be that
the relation between $C_{\varepsilon}$ and $Re_{\lambda}$ found by
\cite{chen2021turbulence} in non-homogeneous conditions is a
reflection of a ``non-homogeneous turbulence cascade'', i.e. a
turbulence cascade which fundamentally concerns non-homogeneous
turbulence? This is the general question which we attempt to address
in the present paper. Previous scale-by-scale analyses in
non-homogeneous conditions include \citeapos{kaneda2020linear} linear
response theory of the inertial sublayer of wall-bounded turbulence
and the study by \citet{afonso2005inhomogeneous} of passive scalars
subjected to non-homogeneous scalar forcings in a homogeneous
gaussian, white-in-time and zero-mean velocity field \citep[see
  also][]{jurvcivsinova2008anomalous}.

The next section lays out a general theoretical framework for the
study of interscale energy transfers which is then applied to the
energy cascade in the special case of homogeneous turbulence. The
Kolmogorov theory's assumptions for obtaining the second order
structure function are spelled out in section 2 for ease of comparison
with the assumptions of section 3 where we develop a theory of energy
transfers for non-homogeneous turbulence and derive predictions for
second order structure functions. In following two sections we
consider the non-homogeneous turbulent wake flows of
\cite{chen2021turbulence} and use their data, which we describe in
section 4, to test in section 5 the structure function predictions of
section 3. We conclude in section 6.

\section{Theoretical framework and energy cascade in homogeneous turbulence}

Turbulence energy cascades consist of interscale exchanges of
turbulent energy which can be formulated in terms of velocity
differences $\delta {\bf u} \equiv {1\over 2}[{\bf u} ({\bf x}^{+},t)
  - {\bf u} ({\bf x}^{-},t)]$ between the fluid velocity ${\bf u}$ at
position ${\bf x}^{+}$ and the fluid velocity ${\bf u}$ at position
${\bf x}^{-}$ at the same time $t$. Following \cite{hill_2001,hill_2002} and using the coordinate transformation ${\bf x}^{+} = {\bf X} +
{\bf r}$, ${\bf x}^{-} = {\bf X} - {\bf r}$, the following equation is
derived from the incompressible Navier-Stokes equation
\begin{equation}
{\partial \over \partial t} \delta {\bf u} + {\bf u}_X\cdot {\bf
  \nabla}_X \delta {\bf u} + \delta {\bf u} \cdot {\bf \nabla}_r
\delta {\bf u} = -{\bf \nabla}_X\delta p + {\nu\over 2} ({\bf
  \nabla}^{2}_X + {\bf \nabla}^{2}_r) \delta {\bf u}
\label{eq1:delta-u}
\end{equation}
where ${\bf u}_X ({\bf X}, {\bf r}, t) \equiv {1\over 2}[{\bf u} ({\bf
    x}^{+},t) + {\bf u} ({\bf x}^{-},t)]$, $\delta p ({\bf X}; {\bf
  r}, t) \equiv {1\over 2}[p ({\bf x}^{+},t) - p ({\bf x}^{-},t)]$ in
terms of the local pressure to density ratio $p$; ${\bf \nabla}_X$ and
${\bf \nabla}^{2}_X$ are the gradient and Laplacian in ${\bf X}$
space; ${\bf \nabla}_r$ and ${\bf \nabla}^{2}_r$ are the gradient and
Laplacian in ${\bf r}$ space. Note that ${\bf X}$ is the centroid
between ${\bf x}^{+}$ and ${\bf x}^{-}$ and that ${\bf r}$ is half the
separation vector between ${\bf x}^{+}$ and ${\bf x}^{-}$. Like ${\bf
  u}_X$ and $\delta p$, the velocity difference $\delta {\bf u}$ can
be expressed as a function of ${\bf X}$ and ${\bf r}$ (and of course
also time $t$).

An energy equation may be obtained from (\ref{eq1:delta-u}) by a
scalar product with $\delta {\bf u}$:
$$
{\partial \over \partial t} \vert \delta {\bf u}\vert^{2} + {\bf
  \nabla}_X (\cdot {\bf u}_X \vert \delta {\bf u}\vert^{2}) + {\bf
  \nabla}_r \cdot (\delta {\bf u} \vert \delta {\bf u}\vert^{2}) =
$$
\begin{equation} 
-2{\bf \nabla}_X \cdot (\delta {\bf u} \delta p) + {\nu\over 2} ({\bf
  \nabla}^{2}_X + {\bf \nabla}^{2}_r) \vert \delta {\bf u}\vert^{2} 
-{1\over 2}(\varepsilon^{+} + \varepsilon^{-})
\label{eq2:delta-u2}
\end{equation}
where use is made of incompressibility and where $\varepsilon^{+}
\equiv \nu ({\partial u_{i} ({\bf x}^{+}, t)\over \partial
  x_{j}^{+}})^{2}$, $\varepsilon^{-} \equiv \nu ({\partial u_{i} ({\bf
    x}^{-}, t)\over \partial x_{j}^{-}})^{2}$ (with summations over
$i,j=1,2,3$). This is the equation which is effectively the basis of
our arguments in the following section but we use it for homogeneous
turbulence in this section for the sake of comparison between
Kolmogorov's theory of homogeneous turbulence and the following
section's theory.

The energy cascade has in fact been mainly and mostly studied for
homogeneous turbulence
\citep{batchelor1953theory,tennekes1972first,pope2000,mathieu2000introduction,
  sagaut2008homogeneous}. Averaging the above equation over
realisations, which is equivalent to averaging over ${\bf X}$ in
homogeneous turbulence, one gets \citep[see][]{frisch1995turbulence}
\begin{equation}
{\partial \over \partial t} \langle\vert \delta {\bf u}\vert^{2}\rangle  + {\bf
  \nabla}_r \cdot \langle\delta {\bf u} \vert \delta {\bf u}\vert^{2}\rangle  =
{\nu\over 2} {\bf \nabla}^{2}_r \langle\vert \delta {\bf u}\vert^{2}\rangle  -
\langle\varepsilon\rangle 
\label{eq3:delta-u2-homog}
\end{equation}
where the brackets signify realisations average and $\langle\varepsilon\rangle 
\equiv \nu \langle({\partial u_{i} ({\bf X}, t)\over \partial X_{j}})^{2}\rangle $.

Equation (\ref{eq3:delta-u2-homog}) shows that the change in time of
the turbulent kinetic energy associated with the separation vector
${\bf r}$ occurs by turbulence dissipation ($-\langle\varepsilon\rangle $ term in
the equation), viscous diffusion in ${\bf r}$ space (${\nu\over 2}
{\bf \nabla}^{2}_r \langle\vert \delta {\bf u}\vert^{2}\rangle $ term) and energy
exchanges between separation vectors (term ${\bf \nabla}_r \cdot
\langle\delta {\bf u} \vert \delta {\bf u}\vert^{2}\rangle $ which is conservative
in ${\bf r}$ space).

To make the link with the energy cascade more apparent, one can follow
\cite{zhou2020energy} and define the average turbulent kinetic energy
in scales smaller than $r=\vert {\bf r}\vert$ as follows: $E(r,t)
\equiv {3\over 4\pi r^{3}}\int_{S(r)} d^{3} {\bf r} \langle\vert
\delta {\bf u}\vert^{2}\rangle $. This is an average over realisations
(equivalently, over ${\bf X}$ in homogeneous turbulence) and over a
sphere of radius $r$ in ${\bf r}$ space. It is easy to verify that
$E=0$ at $r=0$, that $E$ tends to ${1\over 2}\langle\vert {\bf
  u}\vert^{2}\rangle $ as $r\to\infty$ and that $E$ is a monotonically
increasing function of $r$. (An average energy similar to $E$ can be
defined for non-homogeneous flows by integrating over a volume that
has a shape (not necessarily spherical) which takes into account the
shape of the flow and/or the presence of walls. As this volume grows
beyond correlation lengths, this quantity tends to the one-point
kinetic energy averaged over that volume. This is an issue that needs
to be addressed in detail and on its own in a comparative way for
various turbulent flows, and we do so in a dedicated forthcoming
paper.)

The following equation, valid for $r\gg \lambda$, is obtained by
integrating (\ref{eq3:delta-u2-homog}) and using the divergence
theorem:
\begin{equation}
{\partial \over \partial t} E(r,t) + {3\over 4\pi}\int d\Omega_{{\bf
    r}} \langle {\hat{\bf r} \cdot \delta {\bf u}\over r} \vert \delta {\bf
  u}\vert^{2}\rangle  \approx - \langle\varepsilon\rangle 
\label{eq4:Int-delta-u2-homog}
\end{equation}
where $\hat{\bf r}\equiv {\bf r}/r$ and the integral $\int
d\Omega_{{\bf r}}$ is over the solid angle in ${\bf r}$ space. The
validity of this equation is restricted to $r \gg \lambda$ because the
integrated viscous diffusion term has been omitted as it can be shown
to be negligible compared to $\langle\varepsilon\rangle $ for $r \gg
\lambda$ (see Appendix B of \citet{valente2015energy}).

The interscale transfer rate ${3\over 4\pi}\int d\Omega_{{\bf r}}
\langle {\hat{\bf r} \cdot \delta {\bf u}\over r} \vert \delta {\bf
  u}\vert^{2}\rangle $ (scale-space flux if multiplied by $r^3$)
vanishes at $r=0$ and tends to $0$ as $r\to \infty$, as expected from
the concept of an interscale transfer. At a given scale $r$, a
scale-space flux and a cascade from large to small or from small to
large scales corresponds to a negative or positive ${3\over 4\pi}\int
d\Omega_{{\bf r}} \langle {\hat{\bf r} \cdot \delta {\bf u}\over r}
\vert \delta {\bf u}\vert^{2}\rangle $ and contributes a growth or
decrease of $E(r,t)$ in time. A forward cascade on average (from large
to small scales) results from predominance of compression,
i.e. $\hat{\bf r} \cdot \delta {\bf u} < 0$ in ${3\over 4\pi}\int
d\Omega_{{\bf r}} \langle{\hat{\bf r} \cdot \delta {\bf u}\over r}
\vert \delta {\bf u}\vert^{2}\rangle $, whereas an inverse cascade on
average (from small to large scales) results from predominance of
stretching, i.e. $\hat{\bf r} \cdot \delta {\bf u} > 0$ in ${3\over
  4\pi}\int d\Omega_{{\bf r}} \langle {\hat{\bf r} \cdot \delta {\bf
    u}\over r} \vert \delta {\bf u}\vert^{2}\rangle $
\citep{Vassilicos2015}. By virtue of incompressibility, $\int
d\Omega_{{\bf r}} \hat{\bf r} \cdot \delta {\bf u} = 0$: this means
that there are no forward/inverse cascade events without
inverse/forward cascade ones in incompressible turbulence, whether the
cascade is on average predominantly forward or inverse.

The crucial step in Kolmogorov's theory of homogeneous turbulence is
the hypothesis of equilibrium, namely $\vert {\partial \over \partial
  t} E(r,t)\vert \ll \langle\varepsilon\rangle$ for $r$ much smaller
than the integral length-scale $L$ in the limit of very high Reynolds
number. This hypothesis leads to
\begin{equation}
{3\over 4\pi}\int d\Omega_{{\bf r}} \langle {\hat{\bf r} \cdot \delta
  {\bf u}\over r} \vert \delta {\bf u}\vert^{2}\rangle \approx -
\langle\varepsilon\rangle
\label{eq4b:Int-delta-u2-homog-equilib}
\end{equation}
for $\lambda\ll r\ll L$ in the limit where $L/\lambda \to \infty$
(i.e. high Reynolds number limit).

If the Kolmogorov equilibrium hypothesis can be stretched to scales
$r$ as close to $L$ as possible, then
(\ref{eq4b:Int-delta-u2-homog-equilib}) yields the well-known
Taylor-Kolmogorov dissipation scaling
\begin{equation}
\langle\varepsilon\rangle = C_{\varepsilon} k^{3/2}/L .
\label{eq4c:dissipation}
\end{equation}
It is to be noted that this extension of the Kolmogorov equilibrium
hypothesis to scales $r$ that are commensurate with $L$ leads to the
conclusion that $\langle\varepsilon\rangle$ is independent of
viscosity at high enough Reynolds numbers. Whilst it might be possible
to argue the validity of Kolmogorov's hypothesis for very small scales
$r$, it is much harder to do so for $r\sim L$. We must therefore
consider this extension as an additional hypothesis which is
effectively the hypothesis that turbulence dissipation is independent
of viscosity at high enough Reynolds numbers.

The Kolmogorov equilibrium cascade for homogeneous turbulence is a
cascade where the rate $k^{3/2}/L$ of turbulent energy lost by the
largest turbulent eddies into the cascade equals the interscale
transfer rate at all scales $r$ of the inertial range which itself
equals the turbulence dissipation $\langle\varepsilon\rangle$ at the
smallest scales, all this effectively instantaneously. This
demonstrates the central importance of $\langle\varepsilon\rangle$
across all inertial range scales and is the basis of Kolmogorov's two
similarity hypotheses which are therefore expressed in terms of
viscosity $\nu$ and turbulence dissipation $\langle\varepsilon\rangle$
and no other quantity (except inner and outer length-scales bounding
the inertial range).

Kolmogorov's two similarity hypotheses for equilibrium homogeneous
turbulence are the following --quoting from \citet{pope2000}:

{\it Kolmogorov's first similarity hypothesis:} ``At sufficiently high
Reynolds number, the statistics of the small-scale motions ($r\ll L$)
have a universal form that is uniquely determined by $\nu$ and
$\langle\varepsilon\rangle$."

{\it Kolmogorov's second similarity hypothesis:} ``At sufficiently
high Reynolds number, the statistics of the motions of scale $r$ in
the inertial range have a universal form that is uniquely determined
by $\langle\varepsilon\rangle$, independent of $\nu$."

It is then a matter of straightforward dimensional analysis to obtain
\begin{equation}
\langle\vert \delta {\bf u}\vert^{2}\rangle \sim
(\langle\varepsilon\rangle r)^{2/3}
\label{eq4d:K41results}
\end{equation}
in the inertial range of scales $r$.

In the following section we introduce two similarity hypotheses for
non-homogeneous turbulence which replace the highly simplifying
premise of homogeneity and the two Kolmogorov similarity
hypotheses. We also introduce an inner-outer equivalence hypothesis
for turbulence dissipation which replaces Kolmogorov's equilibrium
hypothesis and we use the hypothesis that turbulence dissipation is
independent of viscosity at high enough Reynolds numbers and the
general inter-scale and inter-space energy balance. On this basis, and
with one extra rather weak assumption linking inner to outer scales,
we obtain scalings of second order structure functions in an inertial
range of scales. Our results have similarities with but are also
fundamentally different from (\ref{eq4d:K41results}) and cannot be
obtained by dimensional analysis.

\section{Energy transfers in non-homogeneous turbulence}
\label{sec:theory}
In this section we consider non-homogeneous turbulence where
interscale energy transfers coexist with turbulent energy transport
through physical space and pressure-velocity effects.

Equation (\ref{eq1:delta-u}) is a vector equation. Our analysis in
this section can be carried out effectively unchanged for equation
(\ref{eq2:delta-u2}) but we want to compare our conclusions with
experimental data and so we concentrate on the first component of
equation (\ref{eq1:delta-u}) which, when multiplied by $\delta u_{1}$
(the first component of $\delta {\bf u}$), yields an equation that is
effectively the same as ({\ref{eq2:delta-u2}) except that $\vert
  \delta {\bf u}\vert^{2}$ is replaced by $(\delta u_{1})^{2}$, the
  pressure-velocity term is replaced by $-2 \delta u_{1} {\partial
    \over \partial X_{1}} \delta p$ and $-{1\over 2} (\varepsilon^{+}
  + \varepsilon^{-})$ is replaced by $-{1\over 2} (\varepsilon_{1}^{+}
  + \varepsilon_{1}^{-})$ where $\varepsilon_{1}^{\pm} \equiv \nu
  ({\partial u_{1}({\bf x}^{\pm},t) \over \partial x_{j}})^{2}$ (with
  summation over $j=1,2,3$).
  
We now restrict our attention to turbulent flows, such as those of
\cite{chen2021turbulence}, which are statistically stationary in time at every
location of the flow, and we use the Reynolds decomposition ${\bf
  u}_{X} = \langle {\bf u}_{X}\rangle + {\bf u}_{X}^{\prime}$ by averaging over
realisations or equivalently over time (but not over ${\bf X}$) in
this statistically stationary context. We therefore obtain the
following equation for $\langle(\delta u_{1})^{2}\rangle$:

$$ \langle {\bf u}_X\rangle \cdot {\bf \nabla}_X \langle (\delta u_{1})^{2}\rangle + {\bf
  \nabla}_X \cdot \langle {\bf u}_X^{\prime} (\delta u_{1})^{2}\rangle  + {\bf
  \nabla}_r \cdot \langle \delta {\bf u} (\delta u_{1})^{2}\rangle  
$$
\begin{equation}
   +2\langle \delta u_{1}
{\partial \over \partial X_{1}} \delta p\rangle = {\nu\over 2} ({\bf \nabla}^{2}_X + {\bf \nabla}^{2}_r) \langle (\delta
   u_{1})^{2}\rangle -\varepsilon_{1}
\label{eq5:KHMHdu}
\end{equation}
where $\varepsilon_{1} \equiv {1\over 2}\langle\varepsilon_{1}^{+} +
\varepsilon_{1}^{-}\rangle$ and where use was made of the
incompressibility of $\langle{\bf u}_X\rangle$. This equation is an
energy balance involving mean flow advection of $\langle(\delta
u_{1})^{2}\rangle$ (the term $\langle{\bf u}_X\rangle\cdot {\bf
  \nabla}_X \langle(\delta u_{1})^{2}\rangle$), turbulent transport
rate of $(\delta u_{1})^{2}$ in physical space (the term ${\bf
  \nabla}_X \cdot \langle{\bf u}_X^{\prime} (\delta
u_{1})^{2}\rangle$), interscale energy transfer rate ${\bf \nabla}_r
\cdot \langle\delta {\bf u} (\delta u_{1})^{2}\rangle$, the
velocity-pressure gradient correlation term $2\langle\delta u_{1}
{\partial \over \partial X_{1}} \delta p\rangle$, viscous diffusion of
$\langle(\delta u_{1})^{2}\rangle$ in ${\bf X}$ and ${\bf r}$ spaces,
and turbulence dissipation rate $\varepsilon_{1}$. Note that no
Reynolds decomposition has been applied to $\delta {\bf u}$, similarly
to \citet{klingenberg2020symmetries} who considered two-point moments
of instantaneous velocities, but that such a decomposition can of
course be used provided that this paper's hypotheses and analysis are
suitably reconsidered in the context of knowledge, not currently
available, of the range of applicability of these hypotheses (see
section 5). In the current setting, the term ${\bf \nabla}_X \cdot
\langle{\bf u}_X^{\prime} (\delta u_{1})^{2}\rangle$ includes both the
more traditional turbulent transport and turbulent production by mean
flow. Production may need to be treated separately if
  the validity of some of this section's hypotheses turn out to depend
  on how significant turbulent production is. \citet{kaneda2020linear}
  used the Corrsin length $l_{C}\equiv \langle
  \varepsilon\rangle^{1/2}/S^{3/2}$ to distinguish between scales
  above $l_C$ where the mean shear $S$ is significant and scales below
  $l_C$ where it is not. We use this length-scale $l_C$ in the data
  analysis of section 5 to assess the impact of mean shear on our
  results.

At this point we limit our study to the longitudinal structure
function and therefore take ${\bf r} = (r,0,0)$, but our approach can
also be applied to the transverse structure function,
i.e. $\langle(\delta u_{2})^{2}\rangle$ with ${\bf r} = (r,0,0)$. We
start with our inner/outer similarity hypotheses for non-homogeneous
turbulence which is that there exists a substantial class of turbulent
flows with regions in them where each inter-scale/inter-space
transport process is similar to itself at different locations of the
non-homogeneous turbulent flow as long as it is rescaled with the
appropriate space-local velocity- and length-scales. These velocity-
and length-scales are different depending on whether the scales under
consideration are small enough for viscosity to be a significant
influence (inner) or not (outer). We therefore introduce an outer
length-scale $l_{o} ({\bf X})$ (some integral/correlation length-scale
independent of viscosity) and an inner length-scale $l_{i}({\bf X})$
(dependent on viscosity) such that $l_{i} ({\bf X}) \ll l_{o} ({\bf
  X})$ for high enough Reynolds numbers. The mathematical expression
of our inner and outer similarity hypotheses, which replace
Kolmogorov's statistical homogeneity and two similarity hypotheses in
the present non-homogeneous context, are therefore written for the
two-point terms in equation (\ref{eq5:KHMHdu}) as follows:

\vspace{2mm} 
\noindent
{\it outer similarity for $r\gg l_{i}$:}
\begin{equation}
  \langle (\delta u_{1})^{2}\rangle = V_{o2}^{2}({\bf X}) f_{o2} (r/l_{o})
\label{eq6:SP2o}
\end{equation}
\begin{equation}
{\bf \nabla}_r \cdot \langle \delta {\bf u} (\delta u_{1})^{2}\rangle =
{V_{o3}^{3}({\bf X})\over l_{o}} f_{o3} (r/l_{o})
\label{eq7:SP3o}
\end{equation}
\begin{equation}
{\bf \nabla}_X \cdot \langle {\bf u}_X^{\prime} (\delta u_{1})^{2}\rangle =
{V_{oX}^{3}({\bf X})\over l_{o}} f_{oX} (r/l_{o})
\label{eq8:SPXo}
\end{equation}
\begin{equation}
2\langle \delta u_{1} {\partial \over \partial X_{1}} \delta p\rangle =
{V_{op}^{3}({\bf X})\over l_{o}} f_{op} (r/l_{o})
\label{eq9:SPpo}
\end{equation}
\noindent
{\it inner similarity for $r\ll l_{o}$:}
\begin{equation}
\langle (\delta u_{1})^{2}\rangle = V_{i2}^{2}({\bf X}) f_{i2} (r/l_{i})
\label{eq10:SP2i}
\end{equation}
\begin{equation}
  {\bf \nabla}_r \cdot \langle \delta {\bf u} (\delta u_{1})^{2}\rangle
  = {V_{i3}^{3}({\bf X})\over l_{i}} f_{i3} (r/l_{i})
  \label{eq11:SP3i}
\end{equation}
\begin{equation}
{\bf \nabla}_X \cdot \langle {\bf u}_X^{\prime} (\delta u_{1})^{2}\rangle =
{V_{iX}^{3}({\bf X})\over l_{i}} f_{iX} (r/l_{i})
\label{eq12:SPXi}
\end{equation}
\begin{equation}
2\langle \delta u_{1} {\partial \over \partial X_{1}} \delta p\rangle =
{V_{ip}^{3}({\bf X})\over l_{i}} f_{ip} (r/l_{i})
\label{eq13:SPpi}
\end{equation}
where we have introduced eight velocity scales $V_{o2}$, $V_{o3}$,
$V_{oX}$, $V_{op}$, $V_{i2}$, $V_{i3}$, $V_{iX}$, $V_{ip}$ which are
explicitly dependent on ${\bf X}$ and eight dimensionless functions of
normalised $r$, namely $f_{o2}$, $f_{o3}$, $f_{oX}$, $f_{op}$,
$f_{i2}$, $f_{i3}$, $f_{iX}$, $f_{ip}$ which are not explicitly
dependent on ${\bf X}$. We warn that as one progresses in the
argument, a need appears to modify the inner similarity assumptions
(\ref{eq12:SPXi})-(\ref{eq13:SPpi}). This is explained in the
Appendix.

The turbulence dissipation rate $\varepsilon_{1}$ is also a two-point
statistic but a fundamentally different one because, unlike the ones
listed above, it does not tend to $0$ as $r\to 0$. In fact the
simulations of \cite{alves_portela2017} suggest that it does not
depend significantly on $r$, which is increasingly evidently true as
$r$ becomes smaller. One can of course expect values of $r$ to exist
which are large enough for $\varepsilon_{1}$ to depend on $r$, but
what we effectively assume here is that such high values of $r$ are
beyond our $r$-range of interest which we limit to values of $r$ much
smaller than $l_o$. We therefore define $C_{\varepsilon}$ by
\begin{equation}
  \varepsilon_{1} = C_{\varepsilon}({\bf X}) {V_{o2}^{3}\over l_{o}}
\label{eq14:dissipation}
\end{equation}
where the outer velocity and length scales $V_{o2}$ and $l_{o}$ are
independent of viscosity. In fact it is natural to take $V_{o2}\sim
\sqrt{k}$ and $l_{o}\sim L$, and
we are more precise about the definition of the integral scale $L$
used in this paper's data analysis later in the paper (penultimate
paragraph of section 4). As in Kolmogorov's theory of homogeneous
turbulence, we make the hypothesis that $C_{\varepsilon}$ is
independent of viscosity at high enough Reynolds number.

\subsection{Outer scale-by-scale energy balance}

Injecting
(\ref{eq6:SP2o})-(\ref{eq7:SP3o})-(\ref{eq8:SPXo})-(\ref{eq9:SPpo})
and (\ref{eq14:dissipation}) into (\ref{eq5:KHMHdu}) we are led to
$$ {2l_{o}\over V_{o2}^{2}}(\langle {\bf u}_X\rangle\cdot {\bf \nabla}_X V_{o2})
f_{o2}(r/l_{o}) - (V_{o2}^{-1} \langle {\bf u}_X\rangle\cdot {\bf \nabla}_X l_{o})
{r\over l_{o}} f_{o2}^{\prime}(r/l_{o})
$$
$$ +{V_{oX}^{3}\over V_{o2}^{3}} f_{oX}(r/l_{o}) +{V_{o3}^{3}\over
  V_{o2}^{3}}f_{o3}(r/l_{o}) +{V_{op}^{3}\over
  V_{o2}^{3}}f_{op}(r/l_{o}) =
$$
\begin{equation}
-C_{\varepsilon} + R^{-1} {l_{o}^{2}\over V_{o2}^{2}} [{\bf \nabla}^{2}_X V_{o2}^{2} f_{o2}(r/l_{o})]+ R^{-1}{\bf \nabla}^{2}_{r/l_{0}} f_{o2}(r/l_{o}) 
\label{eq15:SPo-KHMHdu}
\end{equation}
where $R\equiv 2V_{o2}l_{o}/\nu$ is a naturally appearing local (in
${\bf X}$) Reynolds number, ${\bf \nabla}^{2}_{r/l_{0}}$ is the
Laplacian with respect to ${\bf r}/l_{o}$ rather than just ${\bf r}$,
and $f_{o2}^{\prime}(r/l_{o})$ is the derivative of $f_{o2}$ with
respect to its argument $r/l_{o}$.

In the limit where $R\gg 1$, this outer balance simplifies to
$$ {2l_{o}\over V_{o2}^{2}}(\langle {\bf u}_X\rangle \cdot {\bf \nabla}_X V_{o2})
f_{o2}(r/l_{o}) - (V_{o2}^{-1} \langle {\bf u}_X\rangle \cdot {\bf \nabla}_X l_{o})
{r\over l_{o}} f_{o2}^{\prime}(r/l_{o})
$$
\begin{equation}
+{V_{oX}^{3}\over V_{o2}^{3}} f_{oX}(r/l_{o}) +{V_{o3}^{3}\over
  V_{o2}^{3}}f_{o3}(r/l_{o}) +{V_{op}^{3}\over
  V_{o2}^{3}}f_{op}(r/l_{o}) \approx -C_{\varepsilon} ({\bf X})
\label{eq16:SPoRBIG-KHMHdu}
\end{equation}
This is the normalised outer scale-by-scale energy balance and it
involves mean advection (first line in the equation), turbulent
transport in space, interscale turbulence transfer, the
velocity-pressure gradient correlation term and turbulence dissipation
(second line in the equation). There is no viscous diffusion in the
outer scale-by-scale energy balance for $R\gg 1$.

The fact that the right hand side of (\ref{eq16:SPoRBIG-KHMHdu}) is
independent of $r/l_{o}$ implies
$$
{2l_{o}\over V_{o2}^{2}}(\langle {\bf u}_X\rangle \cdot {\bf \nabla}_X V_{o2}) \sim
V_{o2}^{-1} \langle {\bf u}_X\rangle \cdot {\bf \nabla}_X l_{o} 
$$
\begin{equation}
\sim
{V_{oX}^{3}\over V_{o2}^{3}}\sim {V_{o3}^{3}\over V_{o2}^{3}}\sim
{V_{op}^{3}\over V_{o2}^{3}} \sim C_{\varepsilon} ({\bf X})
\label{eq17:outer-result}
\end{equation}
because ${\bf X}$ and $r$ are independent variables. These
proportionalities imply that all four outer velocity scales are
effectively the same, i.e. the same functions of ${\bf X}$, if
$C_{\varepsilon}$ is independent of ${\bf X}$. However, if
$C_{\varepsilon}$ does depend on spatial position ${\bf X}$ as has
been found to be the case in the turbulent wakes of
\cite{chen2021turbulence}, then there are effectively only two
independent outer velocities, $V_{o2}$ and $V_{oX} \sim V_{o3} \sim
V_{op} \sim V_{o2} C_{\varepsilon}^{1/3}$, with two different
dependencies on ${\bf X}$. Furthermore, (\ref{eq17:outer-result})
implies that $V_{oX}$, $V_{o3}$ and $V_{op}$ are independent of
viscosity given our hypothesis on $C_{\varepsilon}$ and $V_{o2} \sim
\sqrt{k}$. The Reynolds number independence of $V_{oX}/V_{o2}$,
$V_{o3}/V_{o2}$ and $V_{op}/V_{o2}$ is used in the last sentence of
section 3.2, and the conclusion (\ref{eq17:outer-result}) of the outer
scale-by-scale balance is used in the Appendix and in sections 3.4 and
3.5.

\subsection{Inner scale-by-scale energy balance}

Injecting
(\ref{eq10:SP2i})-(\ref{eq11:SP3i})-(\ref{eq12:SPXi})-(\ref{eq13:SPpi})
and (\ref{eq14:dissipation}) into (\ref{eq5:KHMHdu}) we are led to
$$ 2V_{i2}(\langle{\bf u}_X\rangle \cdot {\bf \nabla}_X V_{i2}) f_{i2}(r/l_{i}) -
  {V_{i2}^{2}\over l_{i}} (\langle{\bf u}_X\rangle \cdot {\bf \nabla}_X l_{i})
  {r\over l_{i}} f_{i2}^{\prime}(r/l_{i})
$$
$$ +{V_{iX}^{3}\over l_{i}} f_{iX}(r/l_{i}) +{V_{i3}^{3}\over
  l_{i}}f_{i3}(r/l_{i}) +{V_{ip}^{3}\over l_{i}}f_{ip}(r/l_{i}) = 
$$
\begin{equation}
-C_{\varepsilon}{V_{o2}^{3}\over l_{o}} + {\nu\over 2} ({\bf
  \nabla}^{2}_X + {\bf \nabla}^{2}_r) (V_{i2}^{2}f_{i2}(r/l_{i}))
\label{eq18:SPi-KHMHdu}
\end{equation}
where $f_{i2}^{\prime}(r/l_{i})$ is the derivative of $f_{i2}$ with
respect to its argument $r/l_{i}$.

The inner scales depend on viscosity, and we make the assumption that
they are related to the outer scales by the a priori general forms
$V_{i2}^{2} = V_{o2}^{2} g_{2}(R)$, $l_{i}^{2} = l_{o}^{2} g_{l}(R)$,
$V_{i3}^{3} = V_{o3}^{3} g_{3}(R)$, $V_{iX}^{3} = V_{oX}^{3} g_{X}(R)$
and $V_{ip}^{3} = V_{op}^{3} g_{p}(R)$ where all functions $g$
decrease to $0$ as $R\to \infty$. In terms of these relations between
inner and outer scales, equation (\ref{eq18:SPi-KHMHdu}) becomes
$$\sqrt{g_{2}(R)} {2l_{o}\over V_{o2}^{2}}[\langle {\bf
    u}_X\rangle \cdot {\bf \nabla}_X (V_{o2}\sqrt{g_{2}(R)})] f_{i2}(r/l_{i})
$$
$$
- g_{2}(R)[V_{o2}^{-1} g_{l}^{-1/2}(R)\langle {\bf u}_X\rangle \cdot {\bf
    \nabla}_X (l_{o}\sqrt{g_{l}(R)})] {r\over l_{i}} f_{i2}^{\prime}(r/l_{i})
$$
$$+g_{l}^{-1/2}g_{X}{V_{oX}^{3}\over V_{o2}^{3}} f_{iX}(r/l_{i})
+g_{l}^{-1/2}g_{3}{V_{o3}^{3}\over V_{o2}^{3}}f_{i3}(r/l_{i})
$$
$$
+g_{l}^{-1/2}g_{p}{V_{op}^{3}\over V_{o2}^{3}}f_{ip}(r/l_{i}) =
-C_{\varepsilon} + R^{-1}[{l_{o}^{2}\over V_{o2}^{2}}{\bf
    \nabla}^{2}_X (V_{o2}^{2}g_{2} f_{i2})]
$$
\begin{equation}
 + R^{-1}g_{2}g_{l}^{-1}{\bf
  \nabla}^{2}_{r/l_{i}} f_{i2}(r/l_{i})
\label{eq19:SPi-KHMHdu-norma}
\end{equation}
where ${\bf \nabla}^{2}_{r/l_{i}}$ is the Laplacian with respect to
${\bf r}/l_{i}$.

Taking the limit $R\gg 1$, the mean flow advection terms in the first
line of equation (\ref{eq19:SPi-KHMHdu-norma}) tend to $0$ because
$g_{2}(R)\to 0$ and so does the viscous diffusion term in ${\bf X}$
space on the right hand side. In this high Reynolds number limit we
are therefore left with
$$g_{l}^{-1/2}g_{X}{V_{oX}^{3}\over V_{o2}^{3}} f_{iX}(r/l_{i})
+g_{l}^{-1/2}g_{3}{V_{o3}^{3}\over V_{o2}^{3}}f_{i3}(r/l_{i})
+g_{l}^{-1/2}g_{p}{V_{op}^{3}\over V_{o2}^{3}}f_{ip}(r/l_{i})
$$
\begin{equation}
 =
-C_{\varepsilon} + R^{-1} g_{2}g_{l}^{-1}{\bf \nabla}^{2}_{r/l_{i}}
f_{i2}(r/l_{i}) .
\label{eq20:SPiRBIG-KHMHdu}
\end{equation}
Viscous diffusion and dissipation must both be present in the inner
scale-by-scale energy budget as $R\to \infty$ and therefore
\begin{equation}
g_{2}(R)/g_{l}(R) \sim R .
\label{eq21:b1a}
\end{equation}
To ensure that non-linearity is also present in some form in this
inner scale-by-scale budget, and given that $V_{oX}/V_{o2}$,
$V_{o3}/V_{o2}$ and $V_{op}/V_{o2}$ are independent of viscosity,
$g_{X}/\sqrt{g_{l}}$, $g_{3}/\sqrt{g_{l}}$ and $g_{p}/\sqrt{g_{l}}$
must also tend to a finite constant (independent of $R$) or to $0$
(perhaps some of the three but not all three) as $R\to \infty$.


\subsection{Intermediate scaling of the second order structure function}

Both similarity forms (\ref{eq6:SP2o}) and (\ref{eq10:SP2i}) hold in
the intermediate range of scales $l_{i}\ll r \ll l_{o}$, i.e.
\begin{equation}
  \langle (\delta u_{1})^{2}\rangle = V_{o2}^{2}({\bf X}) f_{o2}
  (r/l_{o})=V_{i2}^{2}({\bf X}) f_{i2} (r/l_{i}).
\label{eq27:inout}
\end{equation}
Using $V_{i2}^{2}=V_{o2}^{2}g_{2}(R)$ and
$l_{i}^{2}=l_{o}^{2}g_{l}(R)$, the following outer-inner relation
follows:
\begin{equation}
f_{o2} (r/l_{o})=g_{2}(R) f_{i2} ({r\over l_{o}}) g_{l}^{-1/2}(R))
\label{eq28:inout2}
\end{equation}
which implies ${d\over dR}[g_{2} (R) f_{i2} ({r\over l_{o}}
  g_{l}^{-1/2}(R))] =0$ and therefore ${{d\over dR}g_{2}\over g_{2}} =
- {r\over l_{o}} {f'_{i2}\over f_{i2}} {d\over dR}(g_{l}^{-1/2})$
(where $f'_{i2}$ is the derivative of $f_{i2}$ with respect to its
argument).  Hence ${r\over l_{o}} {f'_{i2}\over f_{i2}}$ is
independent of $r/l_o$, which in turn implies
\begin{equation}
f_{i2} (r/l_{i}) \sim (r/l_{i})^{n}
\label{eq29:scaling}
\end{equation}
for $l_{i}\ll r \ll l_{o}$ with
\begin{equation}
g_{2}(R) g_{l}^{-n/2}(R) = Const
\label{eq30:exp}
\end{equation}
independent of $R$.

We have one exponent $n$ and two functions $g_{2}(R)$ and $g_{l}(R)$
to determine, and two relations between them: (\ref{eq21:b1a}) and
(\ref{eq30:exp}). We therefore need one more relation and for this we
introduce an hypothesis which we term inner-outer equivalence for
turbulence dissipation. This hypothesis replaces in non-homogeneous
turbulence Kolmogorov's equilibrium hypothesis for homogeneous
turbulence.

\subsection{Hypothesis of inner-outer equivalence for turbulence dissipation}

The inner-outer equivalence hypothesis states that the turbulence
dissipation should depend on inner variables in the same way that it
depends on outer variables. Applied to (\ref{eq14:dissipation}), it
states that if the turbulence dissipation rate (per unit mass) is
equal to $C_{\varepsilon} V_{o2}^{3}/l_{o}$ in terms of outer
variables, then it should also be equal to $C_{\varepsilon}
V_{i2}^{3}/l_{i}$ in terms of inner variables. Taking account of
$V_{i2}^{2} = V_{o2}^{2} g_{2}(R)$ and $l_{i}^{2} = l_{o}^{2}
g_{l}(R)$ we obtain a third relation, namely
\begin{equation}
  g_{l}(R) = g_{2}^{3}(R).
\label{eq32new:sim-exp}
\end{equation}
Relations (\ref{eq21:b1a}), (\ref{eq30:exp}) and
(\ref{eq32new:sim-exp}) yield
\begin{equation}
g_{2}(R)\sim R^{-1/2} , g_{l}(R)\sim R^{-3/2}, n=2/3.
\label{eq33:exp}
\end{equation}
We can also use $g_{l}^{-1/2}g_{3} = Const$ stated in the sentence
under (\ref{eq21:b1a}) and also given in (\ref{eq23:cb2}) to obtain
$g_{3}\sim R^{-3/4}$. These are Kolmogorov-looking power laws and
exponents in a non-Kolmogorov situation where the non-homogeneity of
the turbulence is in fact of the essence.

The inner-outer equivalence for turbulence dissipation can be
understood quite broadly and does not strictly rely on
(\ref{eq14:dissipation}). It can also be taken to mean that if we have
$C_{\varepsilon} \sim V_{o3}^{3}/V_{o2}^{3}$ as per
(\ref{eq17:outer-result}), then we must also have $C_{\varepsilon}
\sim V_{i3}^{3}/V_{i2}^{3}$. The ratio between the outer velocity
scales $V_{o3}$ and $V_{o2}$ is therefore the same as the ratio
between the inner velocity scales $V_{i3}$ and $V_{i2}$, i.e.
${V_{o3}\over V_{o2}} = {V_{i3}\over V_{i2}}$. Taking account of
$V_{i3}^{3} = V_{o3}^{3} g_{3}(R)$ and $V_{i2}^{2} =
  V_{o2}^{2} g_{2}(R)$ we obtain
\begin{equation}
g_{2}(R)=g_{3}^{2/3}(R).
\label{eq32:sim-exp}
\end{equation}
Relations (\ref{eq21:b1a}), (\ref{eq23:cb2}), (\ref{eq30:exp}) and
(\ref{eq32:sim-exp}) yield the exact same power laws and exponents
(\ref{eq33:exp}) and $g_{3}\sim R^{-3/4}$.

\subsection{Predictions}

The power laws and exponents (\ref{eq33:exp}) and $g_{3}\sim R^{-3/4}$
obtained by our attempt at a theory of non-homogeneous turbulence (of
course restricted by the yet unknown domain of validity of the
assumptions we based it on) imply the following relations between
inner and outer variables:
\begin{equation}
l_{i} = l_{o} R^{-3/4}
\label{eq34:l}
\end{equation}
\begin{equation}
V_{i2} = V_{o2} R^{-1/4}
\label{eq35:V2}
\end{equation}
\begin{equation}
V_{i3} = V_{o3} R^{-1/4}.
\label{eq36:V3}
\end{equation}
Note that $l_{i}$ is the Kolmogorov length $\eta \equiv
(\nu^{3}/\langle \varepsilon\rangle)^{1/4}$ and $V_{i2}$ is the
Kolmogorov velocity $u_{\eta} \equiv (\langle \varepsilon \rangle
\eta)^{1/3}$ only if $C_{\varepsilon}$ is independent of ${\bf X}$.
However, $l_i$ and $V_{i2}$ are different from $\eta$ and $u_{\eta}$,
respectively, if $C_{\varepsilon}$ varies in physical space, as is
indeed the case in the non-homogeneous turbulence experiments of
\cite{chen2021turbulence}. To be more specific, whilst $l_i$ and
$V_{i2}$ have the same dependencies on viscosity as $\eta$ and
$u_{\eta}$ respectively, they have different dependencies on ${\bf X}$
than $\eta$ and $u_{\eta}$ when the turbulence is non-homogeneous and
$C_{\varepsilon}$ varies with ${\bf X}$.

Note also that $n=2/3$ means $f_{i2} \sim (r/l_{i})^{2/3}$ in the
intermediate range $l_{i}\ll r\ll l_{o}$. Going back to
(\ref{eq10:SP2i}), we can define $f_{i2}^{*} (r/l_{i}) \equiv
(r/l_{i})^{-2/3}f_{i2}(r/l_{i})$ and therefore write
\begin{equation}
\langle (\delta u_{1})^{2}\rangle = V_{i2}^{2} (r/l_{i})^{2/3} f_{i2}^{*} (r/l_{i})
\label{eq37:twothirdslaw}
\end{equation}
for $r\ll l_{o}$. Making use of (\ref{eq34:l}) and (\ref{eq35:V2}),
equation (\ref{eq37:twothirdslaw}) becomes
\begin{equation}
\langle (\delta u_{1})^{2}\rangle = V_{o2}^{2} (r/l_{o})^{2/3} f_{i2}^{*} (r/l_{i})
\sim k (r/L)^{2/3} f_{i2}^{*} (r/l_{i})
\label{eq38:TwoThirdsLaw}
\end{equation}
and like (\ref{eq37:twothirdslaw}), it is expected to be valid for
$r\ll l_o$. In the intermediate range $l_{i}\ll r\ll l_{o}$,
$f_{i2}^{*}$ is expected to be independent of $r/l_{i}$. We made use
in (\ref{eq38:TwoThirdsLaw}) of $ V_{o2}^{2} \sim k$ and $l_{o} \sim
L$.

It must be stressed that our theory's prediction
(\ref{eq38:TwoThirdsLaw}) is different from the Kolmogorov form
$\langle(\delta u_{1})^{2}\rangle = (\langle \varepsilon \rangle
r)^{2/3} f_{i2}^{*} (r/\eta)$ because
the turbulence dissipation does not vary in space as $k^{3/2}/L$, and
$l_{i}$ does not vary in space as $\eta$ either.
Our theory's prediction (\ref{eq38:TwoThirdsLaw}) would have been
identical to Kolmogorov's prediction
if the turbulence was homogeneous and therefore neither turbulence
dissipation nor $k^{3/2}/L$ varied in space.

Our theory also makes a prediction for ${\bf \nabla}_r
  \cdot \langle \delta {\bf u} (\delta u_{1})^{2}\rangle$. Using
  (\ref{eq14:dissipation}) in conjunction with
  (\ref{eq17:outer-result}) which was obtained from the outer
  scale-by-scale energy balance in section 3.1, we get
  $V_{o3}^{3}/l_{o}\sim \varepsilon_{1}$. From (\ref{eq34:l}) and
  (\ref{eq36:V3}), $V_{i3}^{3}/l_{i}\sim V_{o3}^{3}/l_{o}$, hence
  $V_{i3}^{3}/l_{i}\sim \varepsilon_{1}$. The outer and inner
  similarity forms (\ref{eq7:SP3o}) and (\ref{eq11:SP3i}) become ${\bf
    \nabla}_r \cdot \langle \delta {\bf u} (\delta u_{1})^{2}\rangle =
  \varepsilon_{1} f_{o3}(r/l_{o})$ and ${\bf \nabla}_r \cdot \langle
  \delta {\bf u} (\delta u_{1})^{2}\rangle = \varepsilon_{1}
  f_{i3}(r/l_{i})$ respectively, together implying
\begin{equation}
{\bf \nabla}_r \cdot \langle \delta {\bf u} (\delta
  u_{1})^{2}\rangle \sim \varepsilon_{1}
\label{eq3e:asifK41}
\end{equation}
in the intermediate range $l_{i}\ll r \ll l_{o}$. This
  result may form the basis for explaining the observation by
  \citet{alves_portela2017} of orientation-averaged non-linear
  interscale transfer rates approximately equal to minus the
  turbulence dissipation rate over significant ranges of separation
  distances in a turbulent wake's near-field where all the
  inhomogeneity-related energy processes in the scale-by-scale energy
  balance are actually active. The proportionality
  (\ref{eq3e:asifK41}) obtained for non-homogeneous turbulence where
  all inhomogeneity-related energy processes are active does not have
  the same physical foundation as
  (\ref{eq4b:Int-delta-u2-homog-equilib}) for homogeneous equilibrium
  turbulence where they are not.

Before closing the section it is worth mentioning once again that this
section's arguments can also be applied to the transverse structure
function, i.e. $\langle(\delta u_{2})^{2}\rangle$ with ${\bf r} =
(r,0,0)$, with identical results.

\begin{figure}
   \centerline{\includegraphics[width = \textwidth]{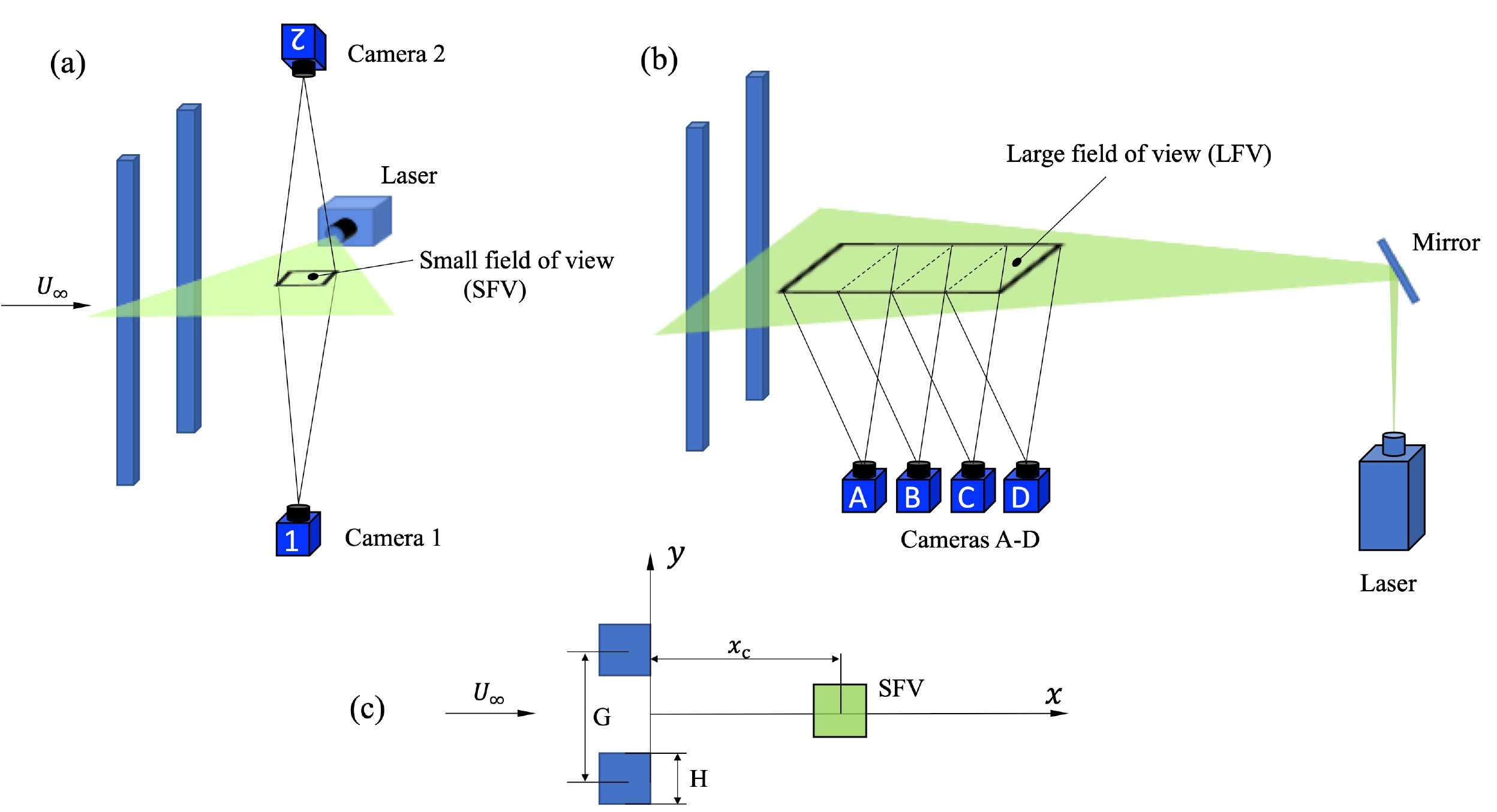}}
  \caption{(a) PIV set-up for the energy dissipation rate
    measurements. (b) PIV set-up for the integral length scale
    measurements. (c) The coordinate system normal to the prisms'
    spanwise direction and definitions of $G$, $H$ and $x_c$. All
    fields of view in (a), (b) and (c) and laser sheets in (a) and (b)
    are in the horizontal $(x,y)$ plane. Figure courtesy of Chen et al
    (2021).}
  \label{fig:exp}
\end{figure}

\section{The experiment of \cite{chen2021turbulence}}
\label{sec:exp}
Given that the laboratory experiments of \cite{chen2021turbulence} provided
the motivation for the developments in the previous section, it is
natural to test the previous section's theory against data from that
experiment. In this section we give a brief reminder of the salient
features of the experiment of \cite{chen2021turbulence} (a detailed
description can of course be found in their paper) and we describe the
data from that experiment which we use in section 5.

\cite{chen2021turbulence} experimented with three different turbulent
wakes of two side-by-side identical square prisms of side length/width
$H=0.03$m. The three different cases corresponded to three different
gap ratios $G/H = 1.25, 2.4, 3.5$ chosen because they give rise to
three qualitatively different flow regimes in terms of dynamics,
large-scale features and inhomogeneity. $G$ is the centre-to-centre
distance between the prisms (see figure \ref{fig:exp} reproduced here
from \cite{chen2021turbulence}). The wind tunnel's test section was 2m
wide by 1m high and the prisms were placed with their spanwise axis
parallel to the tunnel's height. \citet{chen2021turbulence} acquired
data for three incoming velocities $U_\infty = 5, 6, 7.35$ m/s
corresponding to global Reynolds numbers $Re$ $(\equiv U_\infty
H/\nu)$ = 1.0, 1.2 and $1.5 \times 10^4$, respectively.

\begin{figure}
  \centerline{\includegraphics[width = \textwidth]{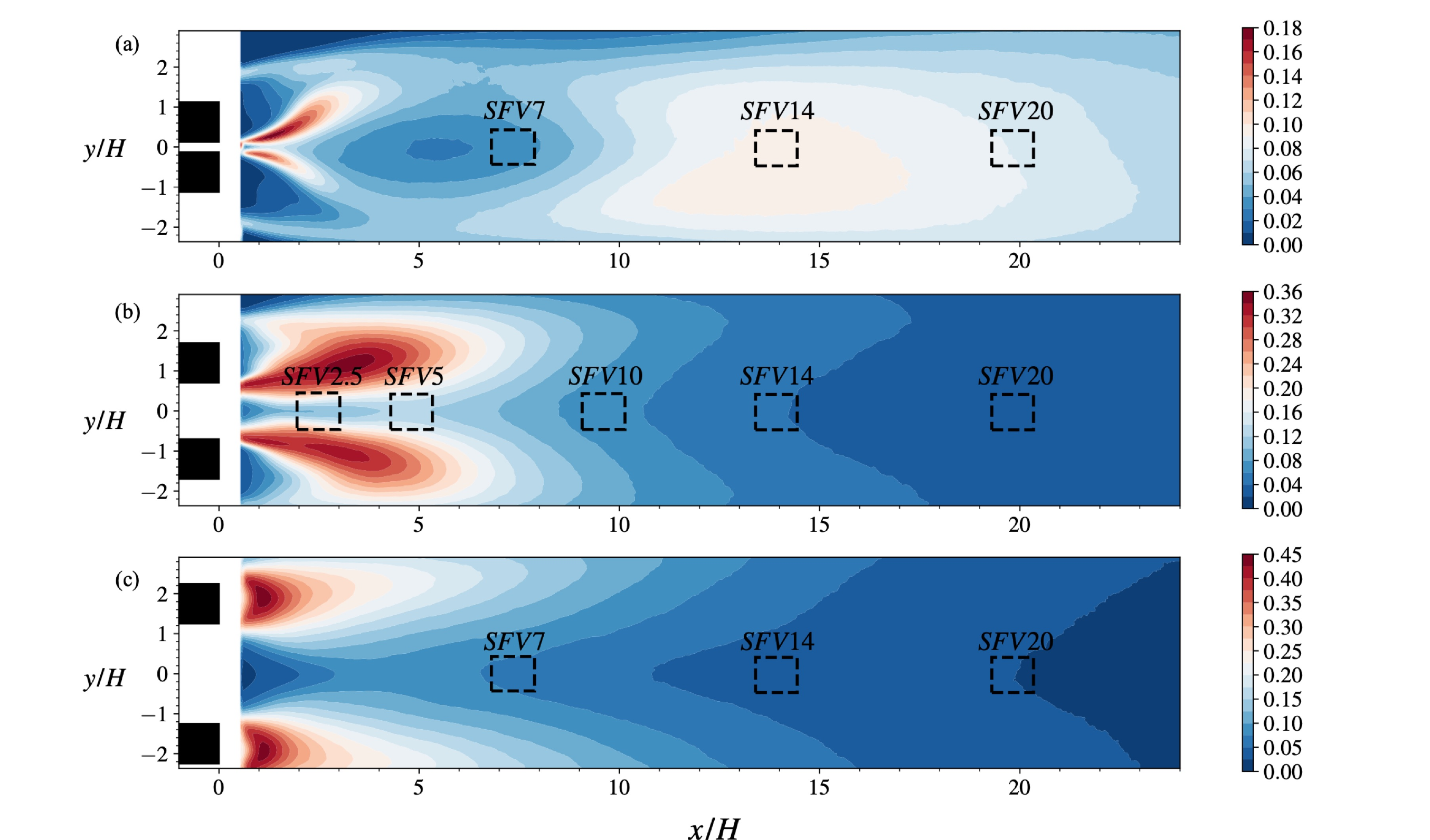}}
  \caption{Spatial distributions of the normalised turbulent kinetic
    energy $k/U_\infty^2$ at Re = $1.0 \times 10^4$ and the positions of the small fields of
    view (dashed squares) for each gap ratio. (a) G/H = 1.25, (b) 2.4,
    and (c) 3.5. Figure courtesy of \cite{chen2021turbulence}.}
  \label{fig:kinetic energy}
\end{figure}

Two different 2D2C PIV set-ups were used, one designed for turbulent
dissipation measurements (see figure \ref{fig:exp}a) and the other for
measurements of integral length scale (see figure \ref{fig:exp}b),
both measuring two horizontal fluid velocity components in the
horizontal $(x,y)$ plane normal to the vertical span of the prisms. A
dual-camera PIV system with small field of view (SFV) was used for the
turbulent kinetic energy dissipation rate (see figure
\ref{fig:exp}a, c). Two sCMOS cameras, one over the top and one under
the bottom of the test section, observed the same SFV and obtained two
independent measurements of the same velocity fields which were then
used to reduce the noise in estimating the energy dissipation rate
(see \cite{chen2021turbulence} for detailed explanations). The SFV
size was similar to the horizontal size of the prisms, specifically
about 1$H$ in streamwise direction by 0.9$H$ in cross-stream direction
(figure \ref{fig:exp}a, c).

\begin{table*}
  \begin{center}
  \caption{Details of the small fields of view (SFV).}
  
  \begin{tabular}[b]{cclllclllcllc}
  \toprule
  $Re$ & & \multicolumn{3}{c}{$1.0\times 10^4$} & & \multicolumn{3}{c}{$1.2\times 10^4$} & &\multicolumn{3}{c}{$1.5\times 10^4$}\\
  \cmidrule{3-5}\cmidrule{7-9}\cmidrule{11-13}
  \vspace{0.1cm} $G/H$ & & 1.25 & 2.4 & 3.5 & & 1.25 & 2.4 & 3.5 &
  & \multicolumn{3}{c}{3.5}\\ \multirow{4}{*}{Cases} & & SFV7 &
  SFV2.5 & SFV7 & & SFV14 & SFV14 & SFV14 & &
  \multicolumn{3}{c}{SFV20}\\ & & SFV14 & SFV5 & SFV14 & & SFV20 &
  SFV20 & SFV20 & &\\ & & SFV20 & SFV10 & SFV20 \\ & & & SFV20
  \\ 
  \bottomrule
  \end{tabular}
  
  \label{tab:SFV}
  \end{center}
\end{table*}
For each gap ratio $G/H$, \citet{chen2021turbulence} took measurements with
SFVs at several downstream positions for two or three global Reynolds
numbers $Re$. The centre of all the SFVs was on the geometric
centerline ($y = 0$), as sketched in figure \ref{fig:exp}c, and
different SFVs differed by different streamwise positions $x_c$ of the
SFV centre.  The measurement cases are summarized in table
\ref{tab:SFV}. The different SFVs are referred to as SFV$\mathcal{N}$
where $\mathcal{N}$ gives an idea in terms of multiples $\mathcal{N}$
of $H$ of the approximate streamwise coordinate $x_c$ of the centre of
the SFV. The positions of the SFVs relative to the prisms can be seen
in figure \ref{fig:kinetic energy}. In this study we use data from all
the cases listed in table \ref{tab:SFV} except $G/H = 1.25$ and $3.5$
at the smallest Reynolds number ($1.0 \times 10^4$).

\begin{figure}
    \centerline{\includegraphics[width =
        \textwidth]{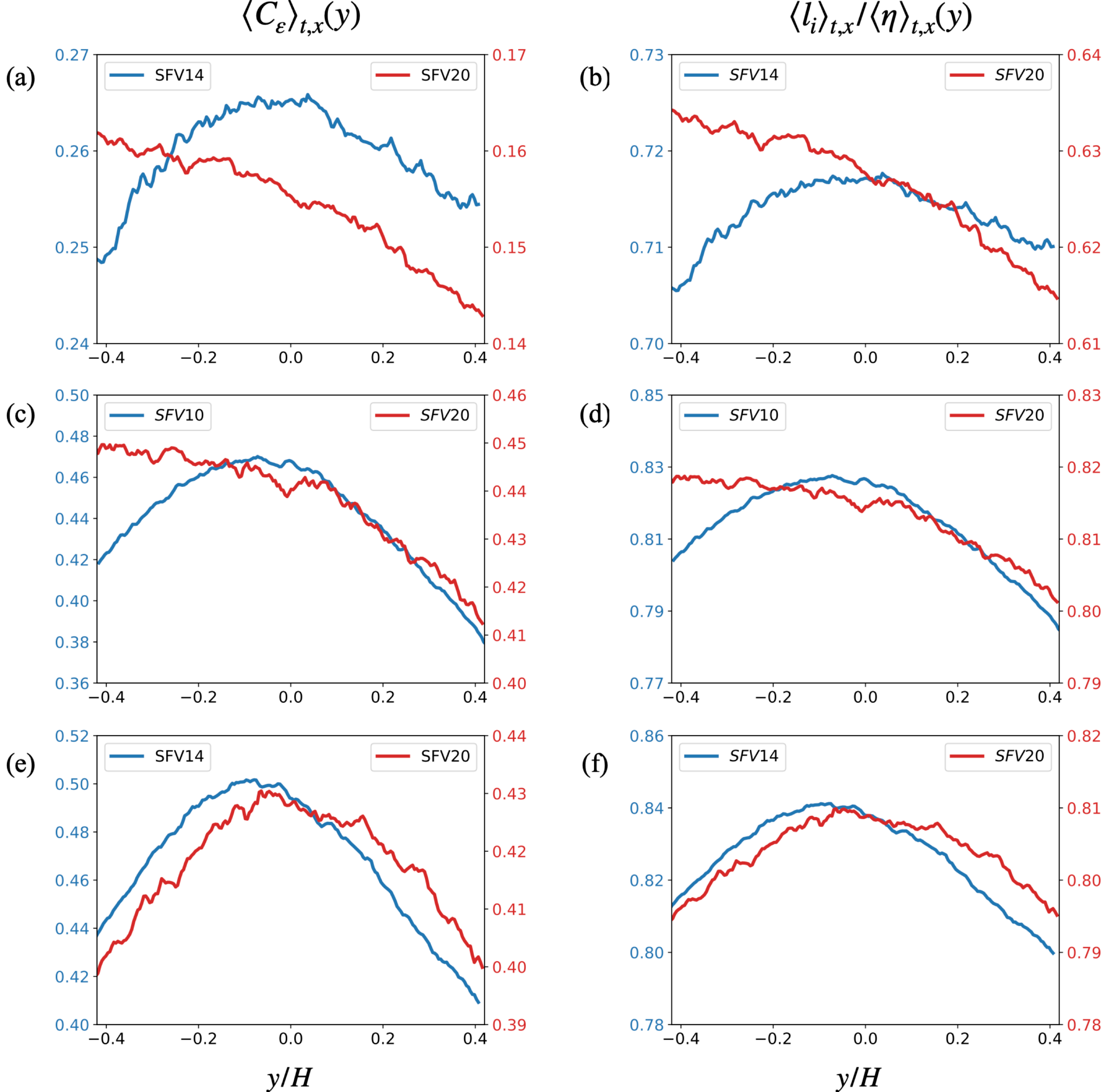}}
    \caption{Left plots (a, c, e) show $\langle
        C_{\varepsilon} \rangle_{t,x}$, where the brackets $\langle ... \rangle_{t,x}$
        signify an average over time $t$ and streamwise coordinate
        $x$, versus $y/H$ within a small field of view. The right
        plots (b, d, f) show $\langle l_{i} \rangle_{t,x}/\langle \eta \rangle_{t,x}$ versus $y/H$,
        where $l_{i}=LR^{-3/4}$ and $\eta = (\nu^{3}/\langle \varepsilon
         \rangle)^{1/4}$. Red lines correspond to SFV20 and blue lines to
        SFV14 (a, b, e, f) or SFV10 (c, d). $Re = 1.0\times 10^4$. Plots
        (a, b) are for $G/H=1.25$; (c, d) for $G/H=2.4$; (e, f) for
        $G/H=3.5$.}
    \label{fig:CepsandLi}
  \end{figure}

The other 2D2C PIV set-up used by \cite{chen2021turbulence} was a
system of four sCMOS cameras arranged consecutively in streamwise
direction to allow integral length scale measurements in a large field
of view (LFV) ranging from $x=0.53H$ to $x=24.3H$ in streamwise
direction and $y=-2.4H$ to $y=2.9H$ in cross-stream direction for $Re
= 1.0 \times 10^4$, and from $x = 11.1H$ to $x=24.9H$ and $y=-2.8H$ to
$y=3H$ for $Re = 1.2$ and $ 1.5 \times 10^4$, see figure
\ref{fig:exp}b.

\begin{figure}
  \centerline{\includegraphics[width = \textwidth]{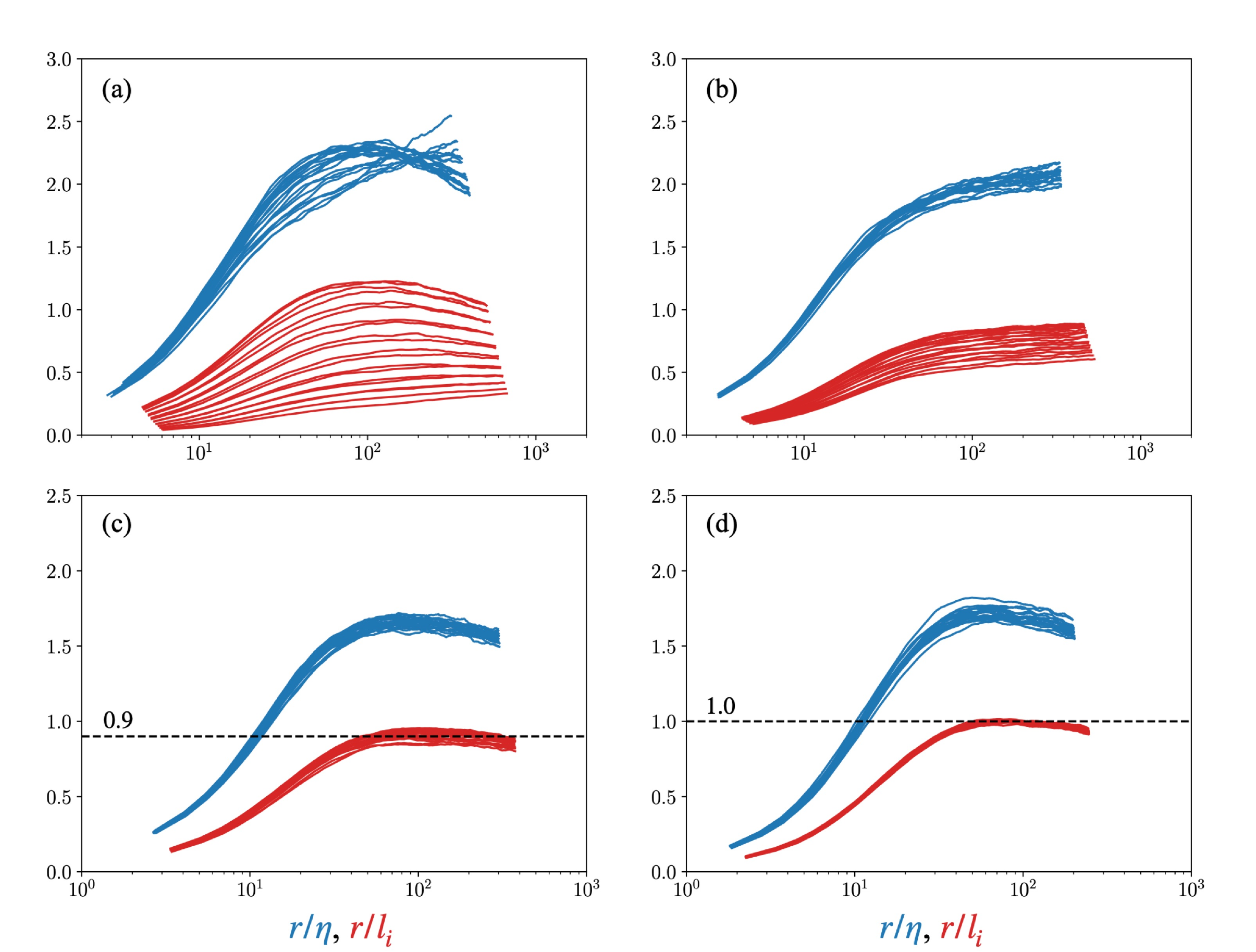}}
  \caption{Comparison between $S_2^u/(\langle \varepsilon \rangle
    r)^{2/3}$ (blue, $r/\eta$ in abscissa where $\eta=(\nu^{3}/\langle
    \varepsilon \rangle )^{1/4}$) and $S_2^u/(k^{3/2}r/L)^{2/3}$ (red,
    $r/l_i$ in abscissa where $l_{i}=LR^{-3/4}$, $R=\sqrt{k}L/\nu$)
    across the flow (different same-colour curves are for different
    values of $y$) in each SFV for G/H = 2.4 at $Re = 1.0 \times
    10^4$. (a) SFV2.5, (b) SFV5, (c) SFV10, (d) SFV20. }
  \label{fig:S2u GH24 LowRe}
\end{figure}

The acquisition frequency was 5Hz for SFV and 4Hz for LFV. 20,000 velocity fields were captured for each measurement,
corresponding to about 67 mins for SFV measurements and 83mins for LFV
measurements.
The final interrogation window size of their PIV analysis was $24
\times 24$ pixels with about $58\%$ overlap, which correspond to a
$312 \mu m$ interrogation window for SFV and 1.6mm for LFV. In the
SFVs, the ratio of the interrogation window size to the Kolmogorov
length scale
varied from 4.5 at the nearest position (SFV2.5) to 2.5 at the
farthest position (SFV20), and was below 3.2 for $x/H >10$ (see figure
2 in \citet{chen2021turbulence}). The turbulent energy dissipation rate was
approximated based on the assumption of local axisymmetry in the
streamwise direction \citep{georgehussein91}. \cite{lefeuvreetal2014}
demonstrated that the turbulent kinetic energy dissipation rate
estimated based on this assumption is a good representation of the
full energy dissipation rate across the stream in the wake of a square
prism, in fact more accurate than the turbulent energy dissipation
rate estimated from the local isotropy assumption.

The turbulent kinetic energy $k$ was estimated from the two horizontal
turbulent velocity components and showed, using direct numerical
simulation data from \cite{zhou2019extreme}, that the ratio of this
estimate to the full turbulent kinetic energy remains about constant
at 0.75 to 0.8 for streamwise distances $x/H \ge 5$. They also
calculated correlation length scales in the streamwise direction of
both streamwise and cross-stream fluctuating velocities. Good
convergence of the auto-correlation functions was achieved for the
cross-stream fluctuations whereas the streamwise fluctuations gave
rise to integral length scales close to $10H$ in those cases when it
was possible to extract values of this length scale from the
auto-correlation function (the auto-correlation function of streamwise
velocity fluctuations often did not converge to $0$). We therefore
adopt their choice of integral length-scale $L$ which is the integral
length scale of the cross-stream turbulence fluctuations in the
streamwise direction. We use their data for $L$ in the following
cases: $G/H=1.25$, $SFV14$ and $SFV20$, where $L/H$ ranges from about
$1.4$ to $1.6$; $G/H=2.4$ where $L/H$ ranges between $0.4$ and $0.5$
in SFV2.5, $0.75$ to $0.8$ in SFV5, and hovers around 1 in SFV14 and
SFV20; and $G/H=3.5$, SFV14 and SFV20, where $L/H$ ranges between
about 1/2 and about 2/3.

In section 5 we use the integral length scale data just mentioned and
also local streamwise and cross-stream velocity data obtained by
\cite{chen2021turbulence} in SFVs as well as local turbulent
dissipation rate $\varepsilon$ and local turbulent kinetic energy data
in SFVs.

\begin{figure}
  \centerline{\includegraphics[width = \textwidth]{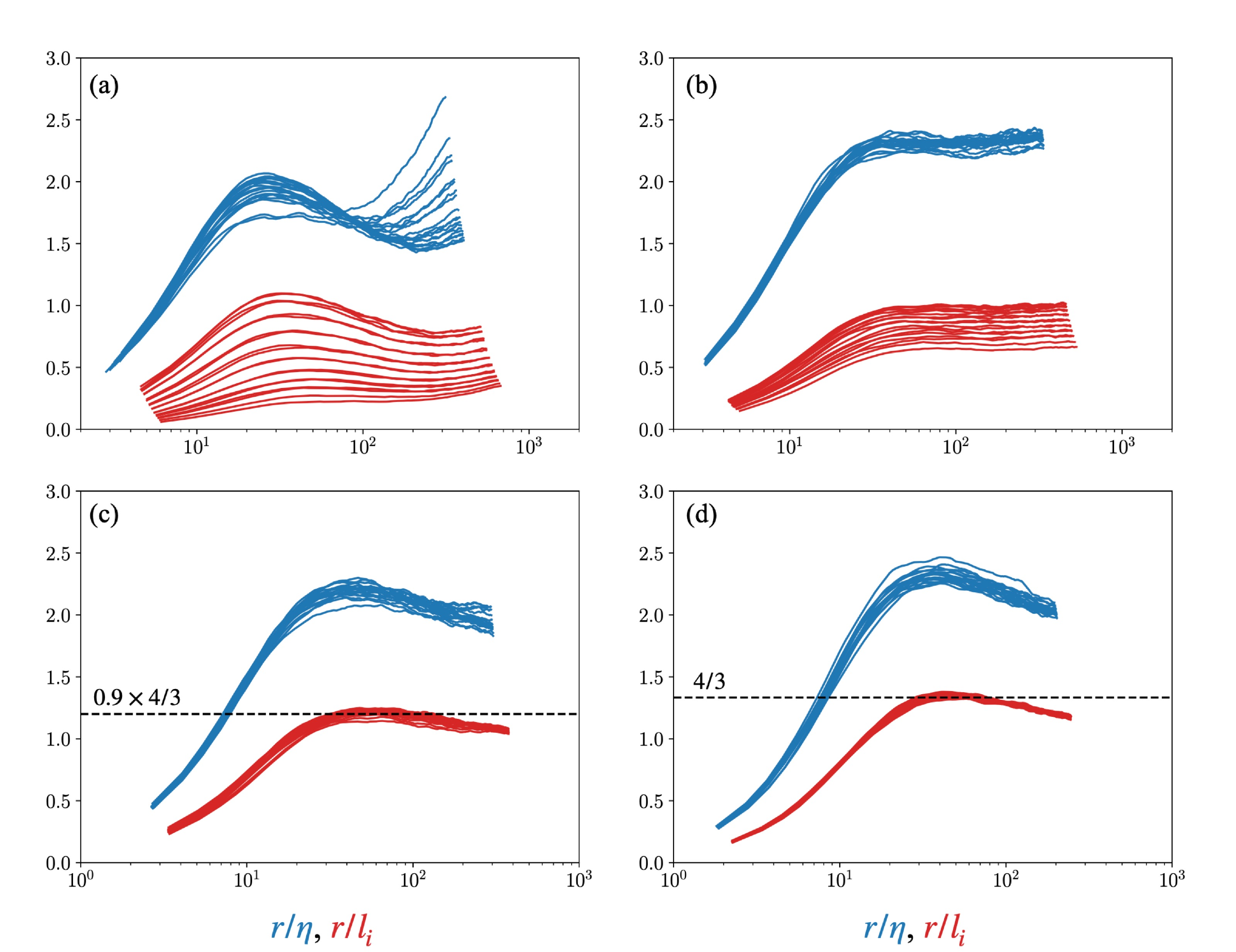}}
  \caption{Comparison between $S_2^v/(\langle \varepsilon \rangle
    r)^{2/3}$ (blue, $r/\eta$ in abscissa where $\eta=(\nu^{3}/\langle
    \varepsilon \rangle)^{1/4}$)) and $S_2^u/(k^{3/2}r/L)^{2/3}$ (red,
    $r/l_i$ in abscissa where $l_{i}=LR^{-3/4}$, $R=\sqrt{k}L/\nu$)
    across the flow (different same-colour curves are for different
    values of $y$) in each SFV for G/H = 2.4 at $Re = 1.0 \times
    10^4$. (a) SFV2.5, (b) SFV5, (c) SFV10, (d) SFV20. }
  \label{fig:S2v GH24 LowRe}
\end{figure}

\section{Scalings of second order structure functions}
In this section, we assess our theory's prediction
(\ref{eq38:TwoThirdsLaw}) for the longitudinal structure function as
well as the equivalent prediction for the transverse structure
function in the three turbulent wake flows described in the previous
section.

Figure \ref{fig:kinetic energy} shows the spatial distribution of the
turbulent kinetic energy in the three flows and illustrates the
qualitative differences between them. Each flow represents one of the
three different flow regimes obtained for different values of $G/H$
\cite[see][]{sumner1999fluid,alam_wake_2011}: the `single-bluff-body
regime' (G/H = 1.25), the `bistable regime' (G/H = 2.4) and the
`coupled vortex regime' (G/H = 3.5). Consistently, the turbulent
kinetic energy field, as well as the mean flow and integral scale
fields (not reproduced here from \cite{chen2021turbulence}),
exhibit distinct spatial distributions and different inhomogeneity
structures which are described and discussed in \cite{chen2021turbulence}.
The inhomogeneity is also present within the SFVs where the turbulence
dissipation coefficient $C_{\varepsilon}$ is found to vary
significantly and systematically with spatial
position in different SFVs (see figure
  \ref{fig:CepsandLi})
in ways that are common in the three different flows,
even though the spatial inhomogeneities of turbulent
  kinetic energy and the turbulence dissipation rate vary from flow to
  flow
(see figure 18 in \cite{chen2021turbulence}). These qualitative
differences in large-scale features and inhomogeneity provide some
variety for the testing of the predictions of section
\ref{sec:theory}.


We use the data of \cite{chen2021turbulence} to calculate the
longitudinal structure function $S_{2}^{u}\equiv <[u'_{1} (x_{0} + r,
  y) - u'_{1}(x_{0}, y)]^{2}>$ and the transverse structure function
$S_{2}^{v}\equiv <[u'_{2} (x_{0} + r, y) - u'_{2}(x_{0}, y)]^{2}>$
where use is made of the Reynolds decomposition $u_{1} = U_{1} +
u'_{1}$ and $u_{2} = U_{2} + u'_{2}$ into mean flow components $U_1$
(streamwise) and $U_{2}$ (cross-stream) and turbulent fluctuating
velocity components $u'_{1}$ and $u'_{2}$. The averaging operation is
over 20,000 velocity field snapshots and we checked that there is no
significant dependence on the choice of streamwise origin $x_{0}$ for
all the $G/H$, $SFV$ and $Re$ cases examined here, except perhaps
$G/H=2.4$ at $SFV2.5$, where changes of $x_{0}$ can create slight
shifts of the curves in figures \ref{fig:S2u GH24 LowRe}a and \ref{fig:S2v GH24 LowRe}a
without significantly changing their shape.

\begin{figure}
    \centerline{\includegraphics[width =
        \textwidth]{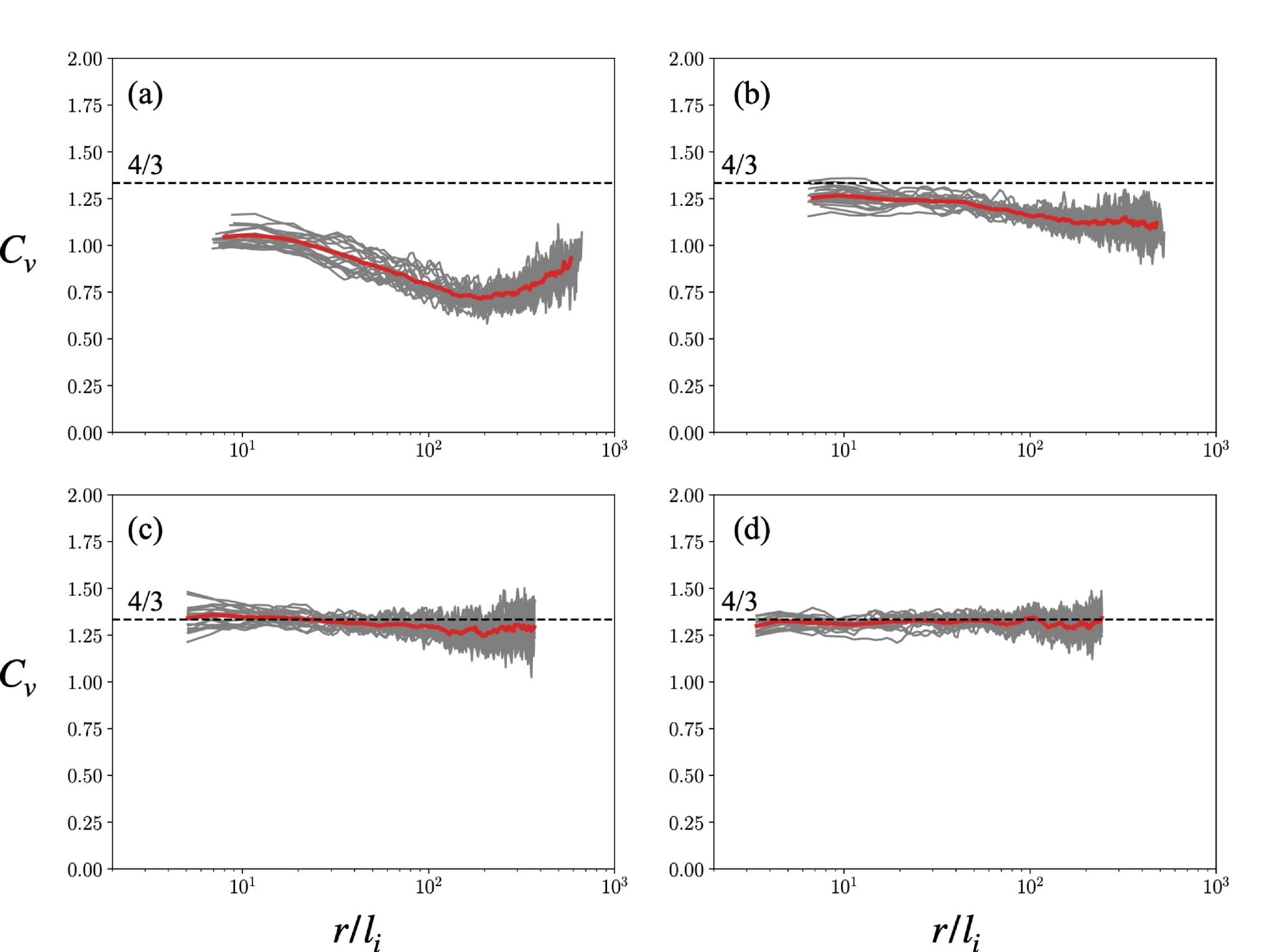}}
    \caption{$C_v (r/l_i)$ versus $r/l_{i}$ (where $l_{i}=LR^{-3/4}$
      with $R=\sqrt{k}L/\nu$) for $G/H = 2.4$ and $Re = 1.0\times
      10^4$ at (a) SFV2.5, (b) SFV5, (c) SFV10 and (d) SFV20. The
      horizontal dashed line is $C_v = 4/3$. Different lines
      correspond to different $y$ and the red line is the average
      curve.}
    \label{fig:fufvRatio GH24}
  \end{figure}
  
The theory of section \ref{sec:theory} was presented for the
longitudinal and transverse structure functions involving $u_1$ and
$u_2$ rather than $u'_1$ and $u'_{2}$. In fact, these structure
functions are insensitive to this difference in all the plots
presented in figures \ref{fig:S2u GH24 LowRe} to \ref{fig:S2v GH all compare} except for two: the plot in figure  \ref{fig:S2u GH24 LowRe}a
which corresponds to the very near field SFV2.5 where the streamwise
mean flow varies appreciably in both the streamwise and the
cross-stream directions within the SFV; and, very slightly, the plot
in figure \ref{fig:S2u GH all}a which is another case where the
streamwise mean flow varies appreciably in the streamwise direction
within the SFV (though less in the cross-stream direction in this
case).

We start the presentation and discussion of our data analysis with
figures \ref{fig:S2u GH24 LowRe} and \ref{fig:S2v GH24 LowRe}
where we plot normalised $S_{2}^{u}$ and $S_{2}^{v}$, respectively, as
functions of normalised $r$ for different cross-stream coordinates $y$
in the case $G/H=2.4$, $Re=10^{4}$. This is the case for which we have
data from four different SFV stations and therefore can get an
impression of dependence on streamwise distance from the pair of
square prisms. Neither the Kolmogorov scalings $S_{2}^{u}/(\varepsilon
r)^{2/3}$ and $S_{2}^{v}/(\varepsilon r)^{2/3}$ versus $r/\eta$ (which
we show for comparison even though they should not be expected to hold
in locally non-homogeneous turbulence) nor our scaling
(\ref{eq38:TwoThirdsLaw}), i.e. $S_{2}^{u}/k(r/L)^{2/3}$ and
$S_{2}^{v}/k(r/L)^{2/3}$ versus $r/l_{i}$, collapse the data well in
the SFV stations closest to the prisms, i.e. SFV2.5 and SFV5. However,
our scaling returns a clearly better collapse than Kolmogorov's in
SFV20 and the two different types of scaling may be judged as
comparable, perhaps with a slight preference for scaling
(\ref{eq38:TwoThirdsLaw}), in SFV10.

It is intriguing that the curves $S_{2}^{v}/k(r/L)^{2/3}$ versus
$r/l_{i}$ appear to plateau at a value that is about $4/3$ the value
where the curves $S_{2}^{u}/k(r/L)^{2/3}$ versus $r/l_{i}$ appear to
plateau in SFV10 and SFV20 (see figures \ref{fig:S2u GH24 LowRe}(c, d) and \ref{fig:S2v GH24 LowRe}(c, d)).
The presence of such a $4/3$ multiplier is well understood in cases
where the turbulence is isotropic and locally homogeneous, in which
cases it is possible to prove the relation $S_{2}^{v} = S_{2}^{u} +
{r\over 2}{\partial S_{2}^{u} \over \partial r}$
\cite[e.g. see][]{pope2000}.  Indeed, if $S_{2}^{u} = A_{u} r^{2/3}$
in a certain range of scales, then $S_{2}^{v} = S_{2}^{u} + {r\over
  2}{\partial S_{2}^{u} \over \partial r}$ implies $S_{2}^{v} =
{4\over 3} A_{u} r^{2/3}$ in that same range of scales.  There is no
local homogeneity in any SFV studied here given that the turbulence is
inhomogeneous within them and that their size is comparable to the
local integral length-scales (between under 1/2 to slightly over 3/2
the integral scale $L$). The usual way to derive $S_{2}^{v} =
S_{2}^{u} + {r\over 2}{\partial S_{2}^{u} \over \partial r}$ can
therefore not be applied here. Nevertheless, our theory's inner
scaling predictions $S_{2}^{u} = k(r/L)^{2/3} f_{u} (r/l_{i})$ and
$S_{2}^{v} = k(r/L)^{2/3} f_{v} (r/l_{i})$, if injected in $S_{2}^{v}
= S_{2}^{u} + {r\over 2}{\partial S_{2}^{u} \over \partial r}$, yield
$f_{v} = {4\over 3} f_{u} + {(r/l_{i})\over
    2}{\partial f_{u} \over \partial (r/l_{i})}$, which means that
  $C_v$ defined as follows should equal $4/3$, i.e.
\begin{equation}
  C_{v} \equiv {f_{v} - {(r/l_{i})\over 2} {\partial f_{u} \over
    \partial (r/l_{i})}\over f_{u}} = 4/3 .
\end{equation}
Note that the value $4/3 = 1 +(2/3)/2$ results from the exponent
$2/3$. If $C_{v}$ differs from $4/3$ then either our theory's
predictions are at fault or $S_{2}^{v} = S_{2}^{u} + {r\over
  2}{\partial S_{2}^{u} \over \partial r}$ does not hold or both.

\begin{figure}
  \centerline{\includegraphics[width = \textwidth]{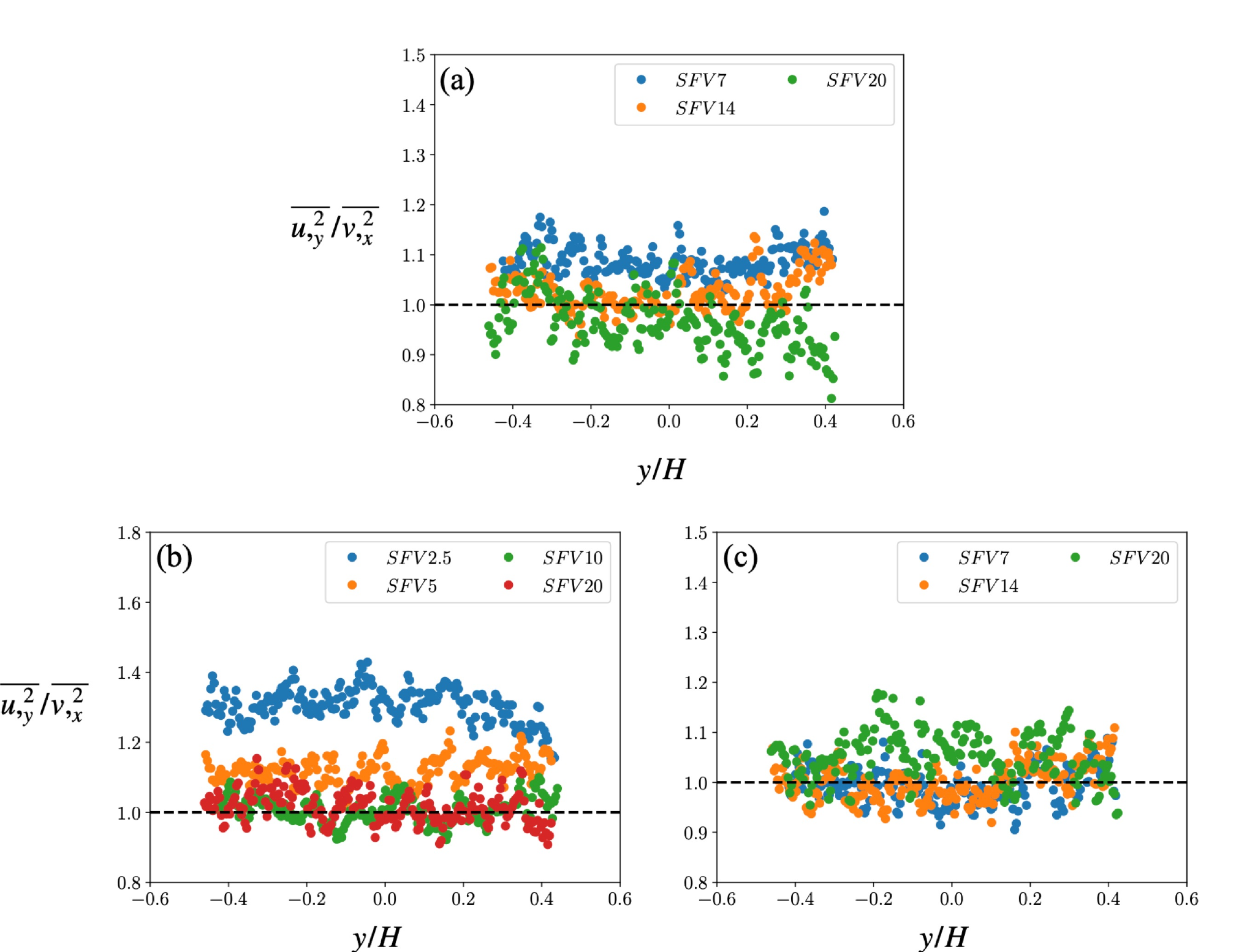}}
  \caption{$\langle u,_y^2\rangle/\langle v,_x^2\rangle$ versus
    normalised cross-stream coordinate $y/H$ in different SFVs for
    $Re=1.0 \times 10^{4}$ and (a) $G/H = 1.25, (b) 2.4, (c) 3.5$.
  }
  \label{fig:isotropy test}
\end{figure}

In figure \ref{fig:fufvRatio GH24} we plot $C_{v}$ versus $r/l_{i}$
obtained from the previous two figures for $G/H=2.5$,
$Re=10^{4}$. $C_{v}$ is clearly well below $4/3$ at all scales in
SFV2.5 where the departure from collapse of $S_{2}^{u}$ and
$S_{2}^{v}$ for different values of $y$ is the greatest. However,
$C_v$ tends to $4/3$ gradually from below as the SFV moves further
away from the prisms, and at SFV20 one can say that $C_v$ is close to
$4/3$ for all values of $y$ over the entire range of $r/l_i$. This is
a non-trivial observation which will require future investigation
because \cite{chen2021turbulence} showed quite clearly that there is
no local homogeneity in SFV20 in terms of turbulent kinetic energy and
turbulent dissipation rate, which means that the usual grounds for
$S_{2}^{v} = S_{2}^{u} + {r\over 2}{\partial S_{2}^{u} \over \partial
  r}$ are absent even though the mean flow may be at its closest to
local homogeneity in SFV20 compared to other SFV stations.  At this
stage, we have no explanation for this result but we do note that it
is consistent with the theory's 2/3 exponent prediction for the second
order structure functions' power law behaviour in the intermediate
range of scales. Figures \ref{fig:S2u GH24 LowRe}(c, d) and \ref{fig:S2v GH24 LowRe}(c, d)
may be giving some partial support to this 2/3 exponent in SFV10 and
SFV20 but over a range of scales that is much smaller than the range
of scales where $C_v$ effectively equals $4/3$ in SFV20 and where
$C_v$ is close to $4/3$ in SFV10.

\begin{figure}
  \centerline{\includegraphics[width = \textwidth]{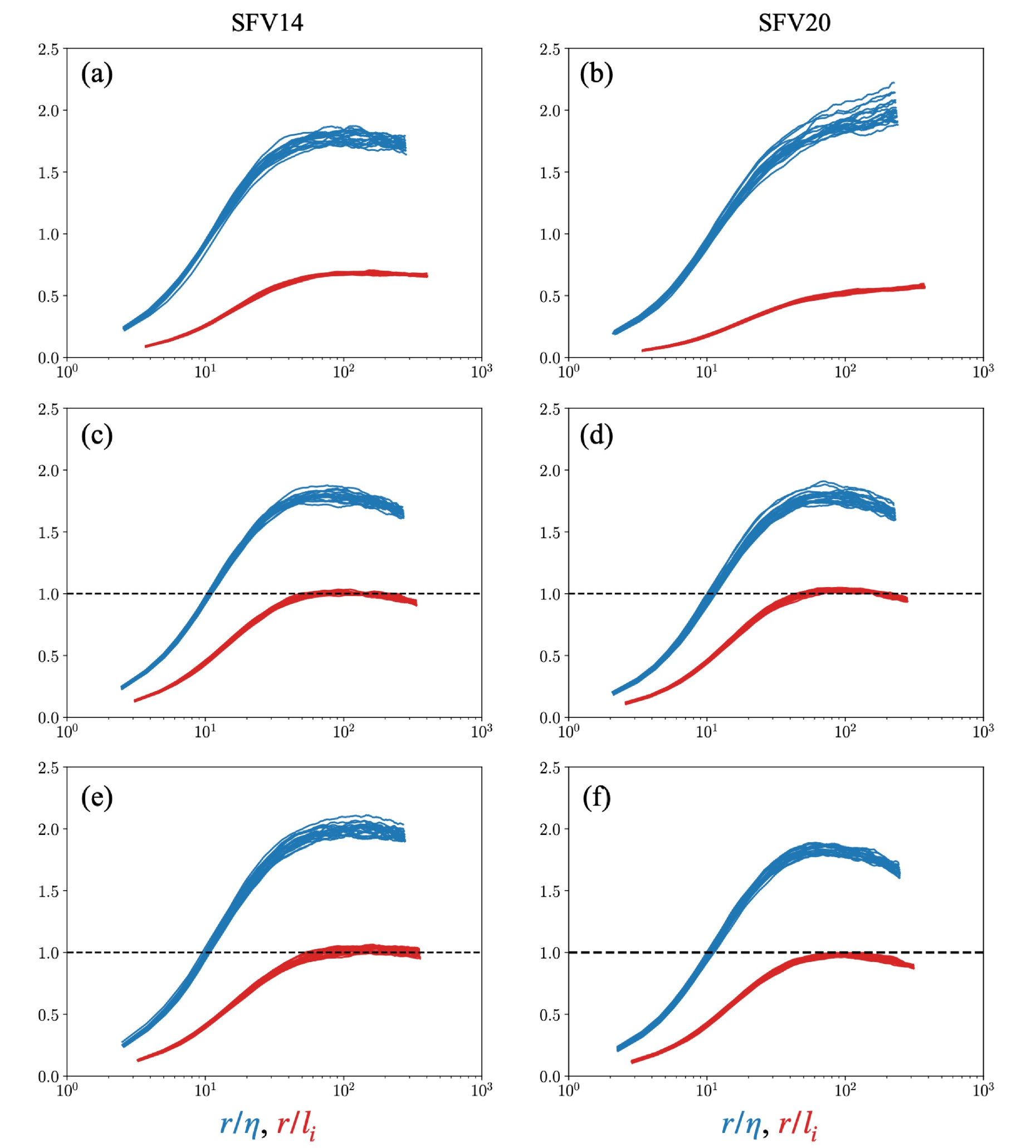}}
  \caption{Comparison between $S_2^u/(\langle\varepsilon\rangle
    r)^{2/3}$ (blue, $r/\eta$ in abscissa where $\eta =
    (\nu^{3}/\langle\varepsilon\rangle)^{1/4}$) and
    $S_2^u/(k^{3/2}r/L)^{2/3}$ (red, $r/l_i$ in abscissa where
    $l_{i}=LR^{-3/4}$, $R=\sqrt{k}L/\nu$) across the SFV14 and SFV20
    for all gap ratios at $Re = 1.2\times 10^4$. Different same-colour
    curves correspond to different $y$ positions. (a) $G/H =1.25$,
    SFV14, (b) $G/H=1.25$, SFV20; (c) $G/H=2.4$, SFV14, (d) $G/H=2.4$,
    SFV20; (e) $G/H=3.5$, SFV14, (f) $G/H=3.5$, SFV20.}
  \label{fig:S2u GH all}
\end{figure}

It may be that $C_v$ is actually more sensitive to anisotropy than
inhomogeneity. We therefore conducted a local isotropy test, reported
in figure \ref{fig:isotropy test}, where we examined the ratio
$\langle u,_y^2\rangle/\langle v,_x^2\rangle$ ($u,_y \equiv \partial
u'_{1}/\partial y$ and $v,_x \equiv \partial u'_{2}/\partial x$)
across the stream (different $y$-positions) in each SFV for different
values of $G/H$. This ratio was taken at the same $x_{0}$ (at a
distance of about $H/10$ from the upstream edge of each SFV) where the
structure functions reported in this paper have been calculated, but
we checked that figure \ref{fig:isotropy test} does not depend significantly on $x_{0}$.
Departures from $\langle u,_y^2\rangle/\langle v,_x^2\rangle = 1$
indicate departures from local isotropy and the biggest such
departures are found at SFV2.5 followed by SFV5 (see figure
\ref{fig:isotropy test}b). The case SFV2.5, $G/H=2.4$ is indeed the
case where our second order structure functions are furthest from
collapse, whether in Kolmogorov variables or along the lines of our
scaling (\ref{eq38:TwoThirdsLaw}) (see figures \ref{fig:S2u GH24 LowRe}a and \ref{fig:S2v GH24 LowRe}a).
There is also unsatisfactory collapse in SFV5 (see figures \ref{fig:S2u GH24 LowRe}b and \ref{fig:S2v GH24 LowRe}b),
and figure $\ref{fig:isotropy test}b$ shows that a significant
departure from local isotropy remains there. In fact, figure
$\ref{fig:isotropy test}$ suggests that $\langle u,_y^2\rangle/\langle
v,_x^2\rangle$ takes values closest to 1 (with a tolerance of about
10\%), and therefore does not indicate significant departures from
local isotropy, further downstream, beyond SFV7 for all three $G/H$
values. Given that figures \ref{fig:fufvRatio GH24} and
\ref{fig:isotropy test} suggest that isotropy may be a prerequisite
for our theory's scalings to hold, a point which will also require
future investigation given that isotropy did not feature explicitly in
our theory's assumptions, we limit the remainder of our data analysis
to SFV14 and SFV20 in all three $G/H$ cases
(\citet{chen2021turbulence} did not take SFV10 measurements for
$G/H=1.25$ and $G/H=3.5$). The data for these SFV stations come with
values of $Re$ higher than $10^4$ which is welcome given that our
theory has been developed for high Reynolds numbers.

\begin{figure}
  \centerline{\includegraphics[width = \textwidth]{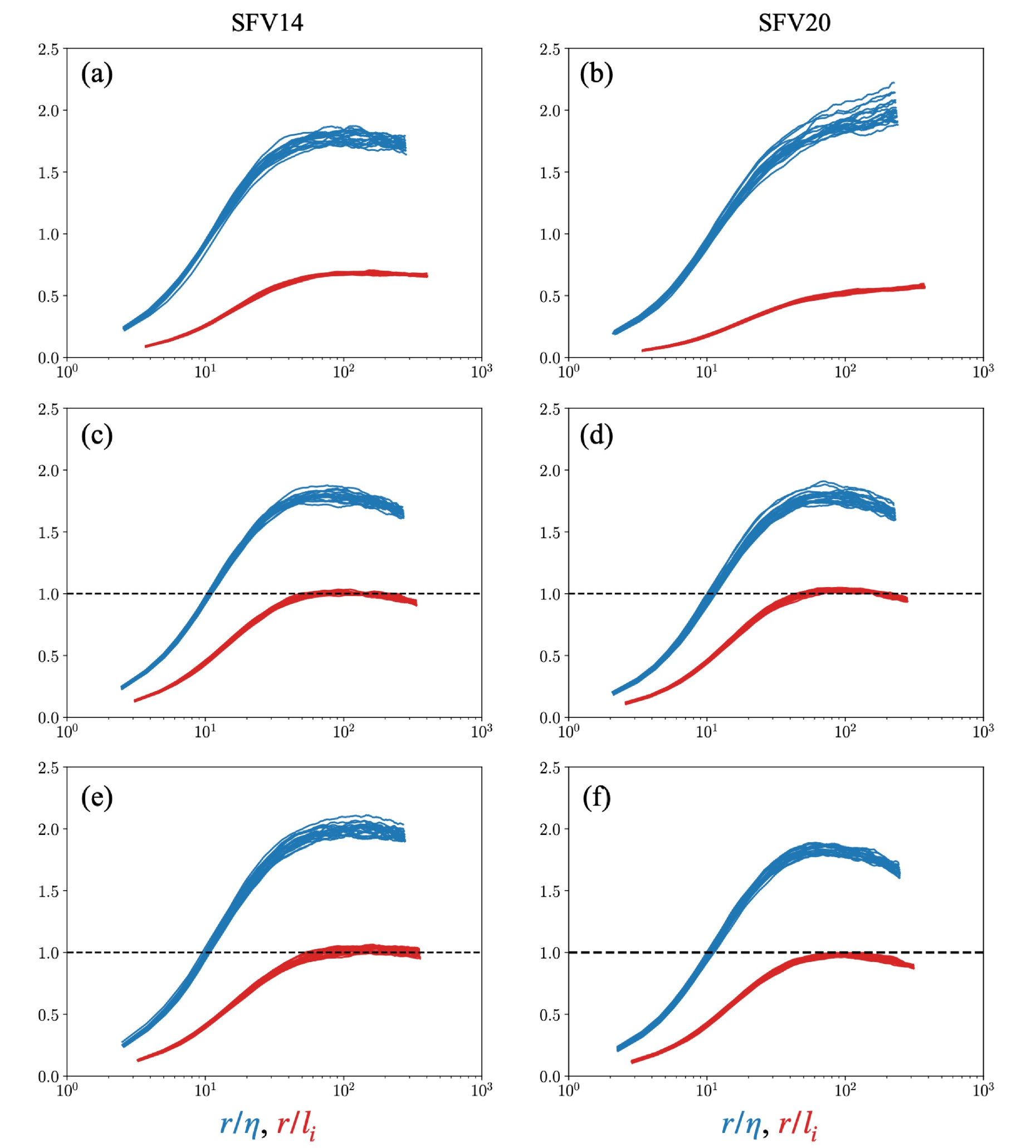}}
  \caption{Comparison between $S_2^v/(\langle\varepsilon\rangle
    r)^{2/3}$ (blue, $r/\eta$ in abscissa where $\eta =
    (\nu^{3}/\langle\varepsilon\rangle)^{1/4}$) and $S_2^v/(k^{3/2}r/L)^{2/3}$ (red,
    $r/l_i$ in abscissa where $l_{i}=LR^{-3/4}$, $R=\sqrt{k}L/\nu$)
    across the SFV14 and SFV20 for all gap ratios at $Re = 1.2\times
    10^4$. Different same-colour curves correspond to different $y$
    positions.  (a) $G/H =1.25$, SFV14, (b) $G/H=1.25$, SFV20; (c)
    $G/H=2.4$, SFV14, (d) $G/H=2.4$, SFV20; (e) $G/H=3.5$, SFV14, (f)
    $G/H=3.5$, SFV20.}
  \label{fig:S2v GH all}
\end{figure}

\begin{figure}
  \centerline{\includegraphics[width = \textwidth]{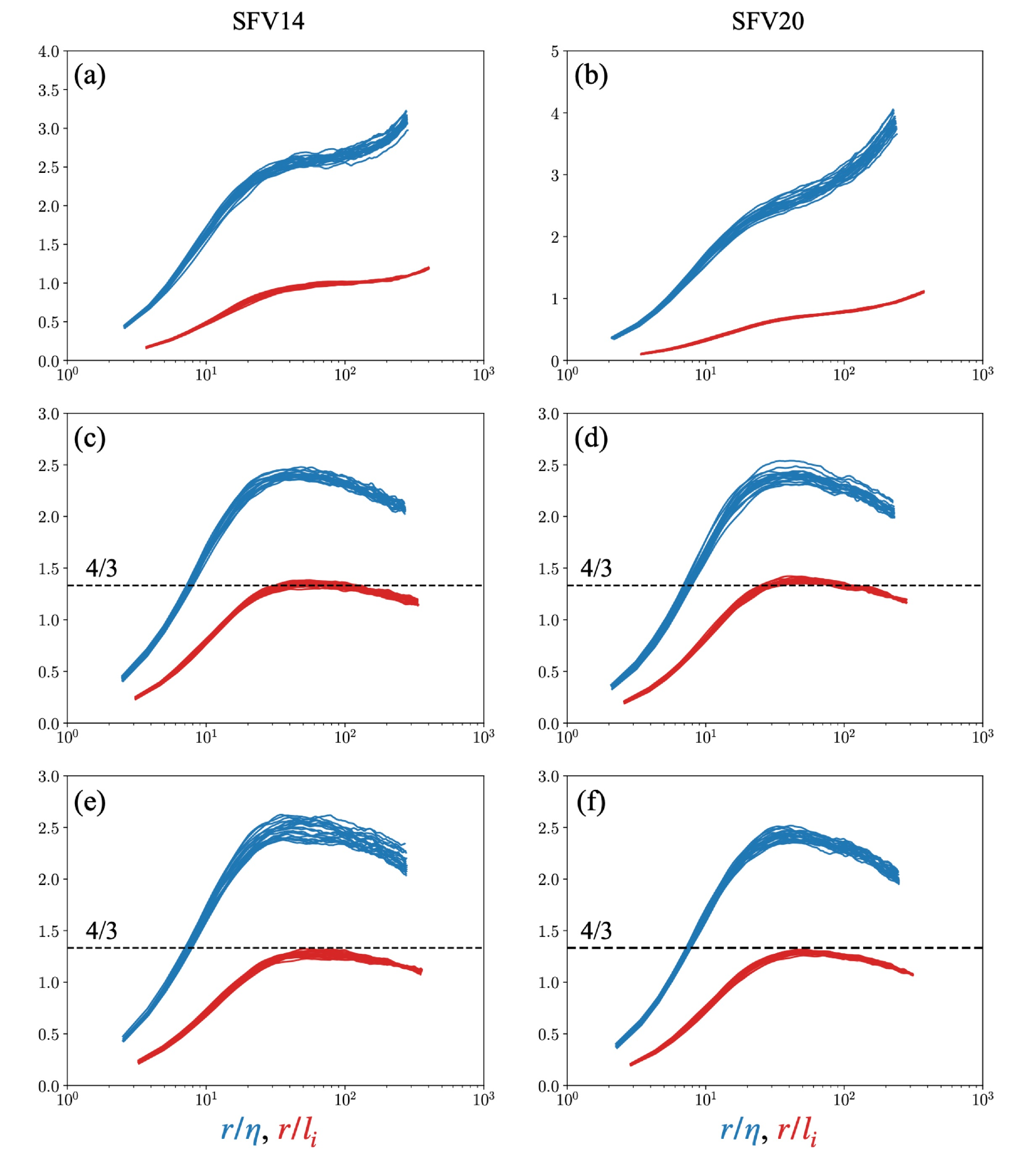}}
  \caption{$C_v (r/l_i)$ versus $r/l_{i}$ (where $l_{i}=LR^{-3/4}$
    with $R=\sqrt{k}L/\nu$) for SFV14 and SFV20 at $Re=1.2\times 10^4$
    for all $G/H$ values. Different curves correspond to different
    $y$-positions and the red line is the average curve. The
    horizontal dashed line in all the plots indicates the value
    $4/3$. (a) $G/H=1.25$, SFV14; (b) $G/H=1.25$, SFV20; (c)
    $G/H=2.4$, SFV14$; (d) $G/H=2.4$, SFV20; (e) $G/H=3.5$, SFV14$;
    (f) $G/H=3.5$, SFV20.}
    \label{fig:fufvRatio all GH}
\end{figure}

In figures \ref{fig:S2u GH all} and \ref{fig:S2v GH all} we plot,
respectively, normalised $S_{2}^{u}$ and $S_{2}^{v}$ as functions of
normalised $r$ for different cross-stream coordinates $y$ and all
three gap ratios, in SFV14 and SFV20 at $Re=12,000$. Our scaling
(\ref{eq38:TwoThirdsLaw}), i.e. $S_{2}^{u}/k(r/L)^{2/3}$ and
$S_{2}^{v}/k(r/L)^{2/3}$ versus $r/l_{i}$, collapses the data well in
both SFV stations and for all three gap ratios. The Kolmogorov
scalings $S_{2}^{u}/(\varepsilon r)^{2/3}$ and $S_{2}^{v}/(\varepsilon
r)^{2/3}$ versus $r/\eta$ (which we show for comparison even though
they should not be expected to hold here) returns a much worse
collapse in all cases. There is a general tendency towards a plateau
in $S_{2}^{u}/k(r/L)^{2/3}$ for $r/l_{i} \ge 40$ which supports the
theory's intermediate range $r^{2/3}$ prediction for the
$r$-dependence of $S_{2}$. However, there is also a difference between
$G/H=1.25$ and the other two $G/H$ values which is most notable in
figure \ref{fig:S2v GH all}: the normalised $S_{2}^{v}$ takes a turn
towards higher values as $r/l_{i}$ increases beyond about 90 in the
$G/H=1.25$ case but not in the other two cases. Consistently with this
observation, $C_v$ is also qualitatively different for $G/H=1.25$ and
for the two other values of $G/H$ (see figure \ref{fig:fufvRatio all
  GH}): it takes values significantly above $4/3$ and in fact
increasing with increasing $r$ for $r/l_{i}$ larger than about $100$
in the $G/H=1.25$ case, whereas nothing of the sort happens in the two
other $G/H$ cases. In fact, $C_v$ is very close to $4/3$ for all $y$
and all $r/l_{i}$ sampled at SFV20 for both $G/H=$ $2.4$ and
$3.5$. The same is the case at SFV14 for $G/H=2.4$ but slightly less
so for $G/H=3.5$ where $C_v$ is close to $4/3$ for $r/l_{i}$ below
about 40 and slightly decreases gradually below $4/3$ with increasing
$r/l_i$ beyond $r/l_{i}=40$.  We note once again that the close to
$4/3$ values of $C_v$ may signify indirect support of the $2/3$
exponent in the power-law dependence on $r$ of the second order
structure functions. A range with such a power law is more visible,
though, in $S_{2}^{u}$ (figure \ref{fig:S2u GH all}) than $S_{2}^{v}$
(figure \ref{fig:S2v GH all}).

\begin{figure}
  \centerline{\includegraphics[width = \textwidth]{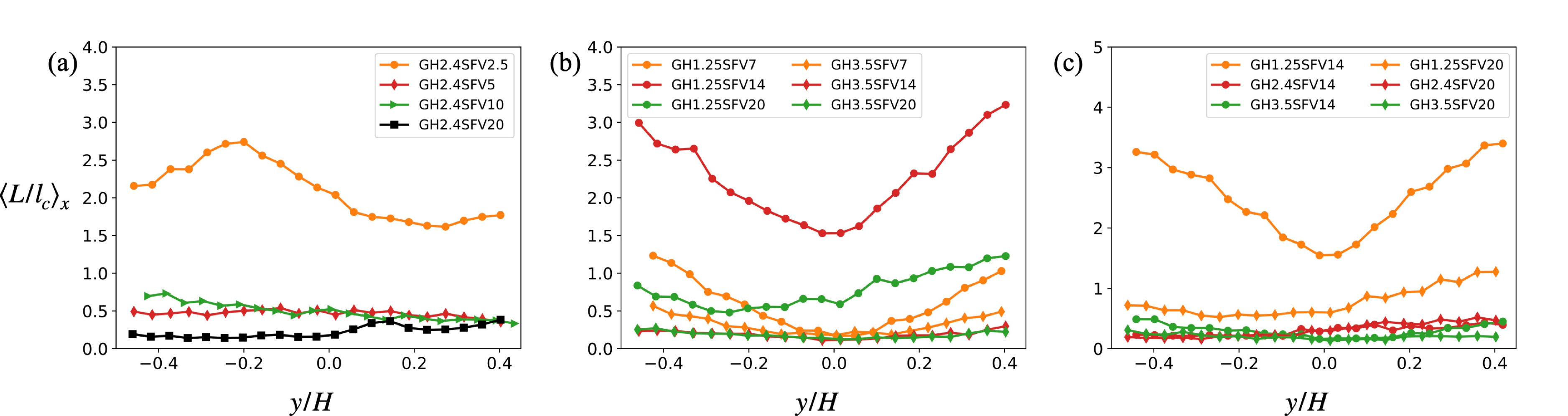}}
  \caption{Cross-stream profiles (along $y$ within an SFV) of $\langle
    L/l_c \rangle_x$, where $\langle \cdot \rangle_x$ denotes
    streamwise averaging within the SFV. (a) $G/H = 2.4$, $Re = 1.0
    \times 10^4$, (b) $G/H = 1.25$ and $3.5$, $Re = 1.0 \times
    10^4$. (c) $G/H = 1.25$, $2.4$ and $3.5$, $Re = 1.2 \times 10^4$.}
  \label{fig:Corrsin}
\end{figure}

\begin{figure}
  \centerline{\includegraphics[width = \textwidth]{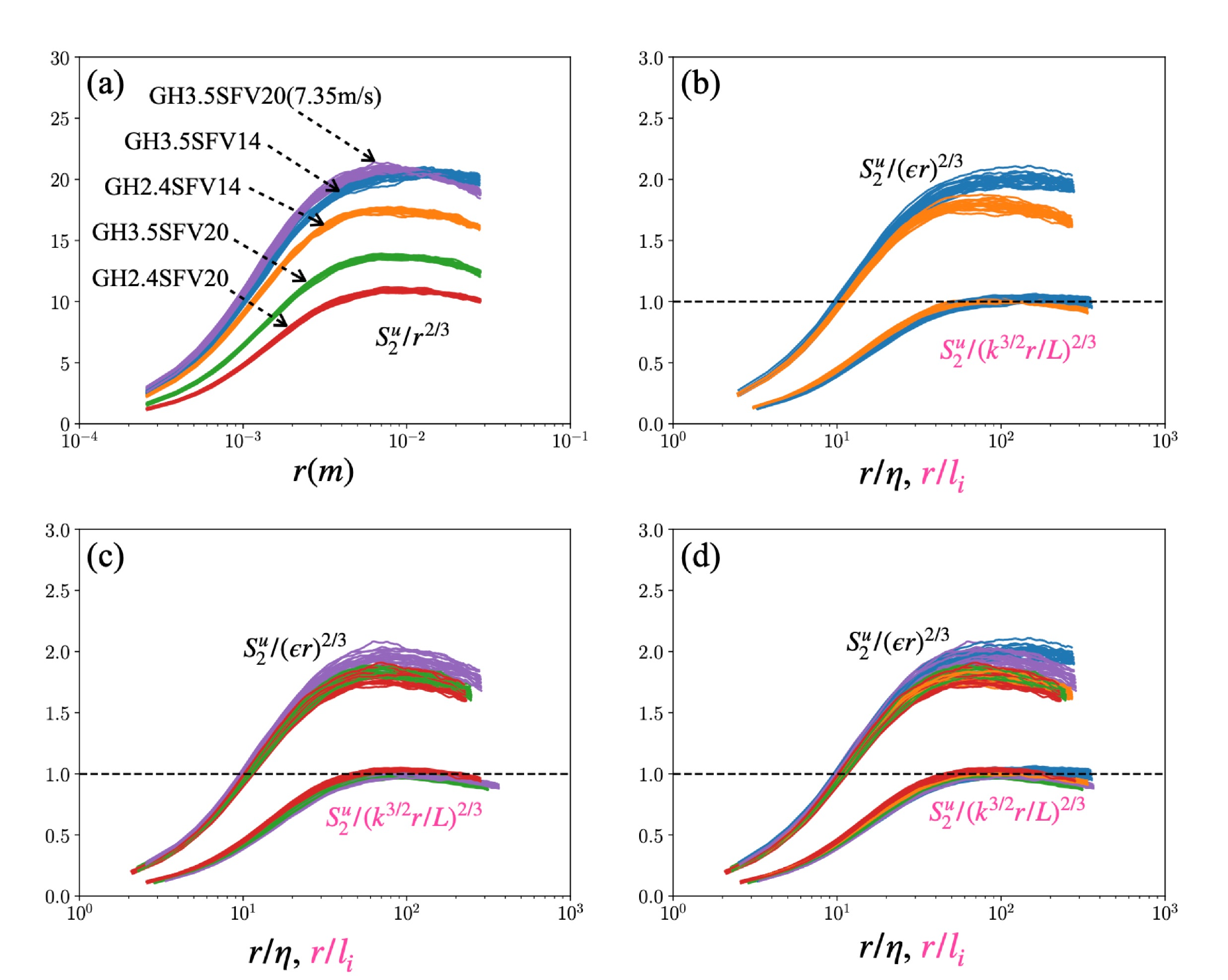}}
  \caption{(a) $S_{2}^{u}/r^{2/3}$ versus $r$ for 5 different cases:
    (i) SFV14, $G/H=2.4$, $Re=1.2\times 10^4$; (ii) SFV14, $G/H=3.5$,
    $Re=1.2\times 10^4$; (iii) SFV20, $G/H=2.4$, $Re=1.2\times 10^4$;
    (iv) SFV20, $G/H=3.5$, $Re=1.2\times 10^4$; (v) SFV20, $G/H=3.5$,
    $R=1.5\times 10^4$. Different same-colour curves correspond to
    different $y$ positions. (b) $S_{2}^{u}/(k^{3/2}r/L)^{2/3}$ versus
    $r/l_{i}$ (where $l_{i}=LR^{-3/4}$ with $R=\sqrt{k}L/\nu$) and
    $S_{2}^{u}/(\langle\varepsilon\rangle r)^{2/3}$ versus $r/\eta$
    (where $\eta = (\nu^{3}/\langle\varepsilon\rangle)^{1/4}$) for
    cases (i) and (ii), i.e. SFV14. (c) Same as (b) but for cases
    (iii), (iv) and (v), i.e. SFV20. (d) Same as (b) and (c) but for
    all cases, SFV14 and SFV20.}
  \label{fig:S2 GH all compare}
\end{figure}

The departures of $C_v$ from $4/3$ in the near field positions SFV2.5
and SFV5 of the $G/H=2.4$ flow (figure \ref{fig:fufvRatio GH24}) could
be accounted for by the departures from local isotropy evidenced in
figure \ref{fig:isotropy test}. However, the departures of $C_v$ from
$4/3$ seen in figure \ref{fig:fufvRatio all GH} for the $G/H=1.25$
flow case cannot be explained that way (see figure \ref{fig:isotropy
  test}). We therefore investigate the possibility that mean shear may
be responsible for these departures. We estimate a
  Corrsin length $l_{C} \equiv \sqrt{\langle \epsilon\rangle /S^{3}}$
  (see \citet{sagaut2008homogeneous}), following
  \citet{kaneda2020linear}, and compare it to our estimate of the
integral length-scale $L$ (see penultimate paragraph of section 4). In
figure \ref{fig:Corrsin} we plot $\langle L/l_{C}\rangle_{x}$ as a
function of cross-stream coordinate $y$ where $\langle \cdot
\rangle_{x}$ symbolises averaging over streamwise coordinate $x$
within a small field of view (SFV) and where $l_C$ is calculated by
taking $S = \vert {\partial U_{1}\over \partial x}\vert +
\vert{\partial U_{1}\over \partial y}\vert + \vert{\partial U_{2}\over
  \partial x}\vert +\vert{\partial U_{2}\over \partial y}\vert$ to
avoid over-estimating $l_C$. For SFV14 and SFV20 in the flow cases
$G/H=2.4$ and $3.5$ where our theory agrees well with the experimental
data, $\langle L/l_{C}\rangle_{x}$ is well below $0.5$, suggesting
that mean shear is absent at the length-scales considered for the
second order structure functions. This agrees with our observation in
this section's fourth paragraph that these structure functions are
effectively the same at $SFV14$ and $SFV20$ for $G/H=2.4$ and $3.5$ if
calculated for the instantaneous or the fluctuating
velocities. Consistently perhaps, $\langle L/l_{C}\rangle_{x}$ is
larger than $1.5$ for SFV2.5, $G/H=2.4$ where neither Kolmogorov's nor
our scalings work (figures  \ref{fig:S2u GH24 LowRe} and \ref{fig:S2v GH24 LowRe}). Turbulence production and
anisotropy should be taken somehow explicitely into account in this
case, which the theory in section 3 does not. Such an extension of our
theory may also be needed for SFV5, $G/H=2.4$ where our theory's
scalings also do not work well (figures \ref{fig:S2u GH24 LowRe} and \ref{fig:S2v GH24 LowRe}) and $\langle
L/l_{C}\rangle_{x}$ takes values up to about 0.75.

However, $\langle L/l_{C}\rangle_{x}$ cannot be, on its own, a
reliable criterion for the applicability of the theory in section
3. It takes values at SFV10, $G/H=2.4$ which are comparable to those
that it takes at SFV5, $G/H=2.4$, yet our theory's collapse for the
second order structure functions is not as bad at SFV10 as it is at
SFV5 (see figures \ref{fig:S2u GH24 LowRe} and \ref{fig:S2v GH24 LowRe}). More dramatically, $\langle
L/l_{C}\rangle_{x}$ is larger than 1.5 at SFV14 in the $G/H=1.25$ flow
case, yet the collapse there is good (figures \ref{fig:S2u GH all}a and \ref{fig:S2v GH all}a) with the only
exception that the exponent $2/3$ may not be present in
$S_{2}^{v}$. At SFV20 in the same flow, $\langle L/l_{C}\rangle_{x}$
takes values between $0.5$ and $1.1$, larger than the values it takes
at SFV14 and SFV20 for $G/H=2.4$ and $3.5$ where the theory works
rather well (figures \ref{fig:S2u GH all} and \ref{fig:S2v GH all}), but well below $1.5$, the smallest
value it takes at SFV14 for $G/H=1.25$. Yet, at SFV20 of this
$G/H=1.25$ flow, the exponent $2/3$ appears absent from both
$S_{2}^{u}$ and $S_{2}^{v}$ even though our scalings collapse them
both very well. Also, $C_v$ is similar at SFV14 and SFV20 of this flow
case (figures \ref{fig:fufvRatio all GH}a, b). All in all, we cannot quite conclude that the
Corrsin length $l_C$ can be used on its own as the basis for an
applicability criterion of our theory (e.g. that the theory appplies
where $\langle L/l_{C}\rangle_{x}$ is smaller than 0.5 but does not
apply where it is larger than 1).
Clearly, the determination of the range and criteria of applicability
of the theory requires more research which we must leave for future
studies.

We now go one step further and explore the possibility that the
scaling (\ref{eq38:TwoThirdsLaw}) may be able to collapse second order
structure functions across flows (i.e. different values of $G/H$), at
different streamwise stations in these flows and for different
Reynolds numbers. Even though we consider only two different SFV
stations (SFV14 and SFV20), only two Reynolds numbers ($Re=12,000$ and
$15,000$, see table \ref{tab:SFV}) and only two of our three flows
($G/H=2.4$ and $3.5$), it does remain worthwhile to test for such
collapse. The $G/H=1.25$ flow case is not included in this test
because of the clear differences between $G/H=1.25$ on the one hand
and $G/H=2.4$, $3.5$ on the other hand which are evidenced in figures
\ref{fig:S2u GH all} and \ref{fig:S2v GH all}.

In figures \ref{fig:S2 GH all compare}a and \ref{fig:S2v GH all
  compare}a we plot, respectively, $S_{2}^{u}/r^{2/3}$ and
$S_{2}^{v}/r^{2/3}$ versus $r$ (in meters) for a wide range of $y$
positions (within the SFV) and five different cases: (i) SFV14,
$G/H=2.4$, $Re=12,000$; (ii) SFV14, $G/H=3.5$, $Re=12,000$; (iii)
SFV20, $G/H=2.4$, $Re=12,000$; (iv) SFV20, $G/H=3.5$, $Re=12,000$; (v)
SFV20, $G/H=3.5$, $R=15,000$. There is no collapse. As a second step,
we plot in figures \ref{fig:S2 GH all compare}b and \ref{fig:S2v GH
  all compare}b, $S_{2}^{u}/(k^{3/2}r/L)^{2/3}$ and
$S_{2}^{v}/(k^{3/2}r/L)^{2/3}$ respectively, both versus $r/l_{i}$,
for cases (i) and (ii) which are for SFV14. The collapse is defensible
but better for $S_{2}^{u}$ than for $S_{2}^{v}$. These two structure
functions are also plotted with Kolmogorov scalings in these two
figures for comparison and it is clear that the Kolmogorov scalings
cannot achieve collapse, particularly for $S_{2}^{u}$. Our third step
is to test for collapse in SFV20. For this, we plot
$S_{2}^{u}/(k^{3/2}r/L)^{2/3}$ and $S_{2}^{v}/(k^{3/2}r/L)^{2/3}$
versus $r/l_{i}$ for cases (iii), (iv) and (v) which are for SFV20
(see figures \ref{fig:S2 GH all compare}c and \ref{fig:S2v GH all
  compare}c). Conclusions are similar. In a final step, we plot all
cases (i), (ii), (iii), (iv) and (v) together in figures \ref{fig:S2
  GH all compare}d and \ref{fig:S2v GH all compare}d. There is a
defensible collapse for $S_{2}^{u}$ across these two flows, two flow
stations, and two Reynolds numbers, much better than if a collapse was
attempted in Kolmogorov variables. One may say that there is also a
more or less acceptable collapse for $S_{2}^{v}$, but it is less sharp
than the collapse of $S_{2}^{u}$. However, a careful look at the
plots, in particular figures \ref{fig:S2v GH all compare} (b, c), reveals
that there remains some dependence on $G/H$ at SFV14 and SFV20,
i.e. some dependence of the similarity function $f_{v}(r/l_{i}, G/H)$
on inlet conditions at these stations.

\section{Conclusions and new research directions}

We developed a theory of non-homogeneous turbulence which we applied
to free turbulent shear flows. The theory predicts that second order
structure functions scale according to equation
(\ref{eq38:TwoThirdsLaw}) in non-homogeneous turbulence, i.e.
\begin{equation}
  S_{2}^{u} = k(r/L)^{2/3} f_{u} (r/l_{i})
\label{eq39:END1}
\end{equation}
and
\begin{equation}
    S_{2}^{v} = k(r/L)^{2/3} f_{v} (r/l_{i})
\label{eq40:END2}
\end{equation}
where $l_{i}$ is an inner length-scale which differs from but has the
same dependence on viscosity as the Kolmogorov length-scale
$\eta$. These scalings take non-homogeneity explicitely into account
and are different from Kolmogorov scalings. However they become
identical to Kolmogorov scalings in homogeneous stationary turbulence.
Note that the theory also leads to the intermediate
  range proportionality (\ref{eq3e:asifK41}) irrespective of the
  presence of inhomogeneity-related energy transfer mechanisms.

\begin{figure}
  \centerline{\includegraphics[width = \textwidth]{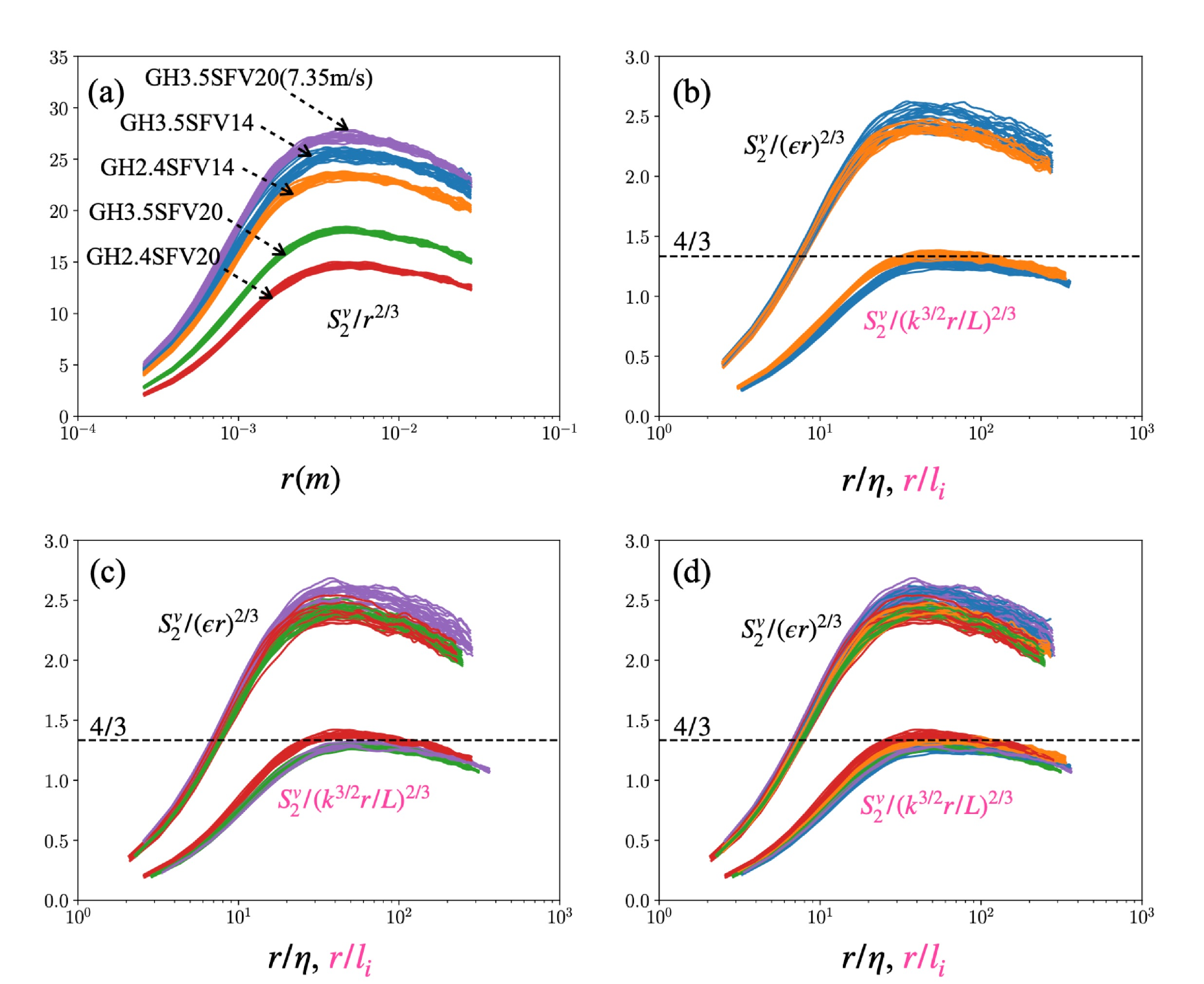}}
    \caption{(a) $S_{2}^{v}/r^{2/3}$ versus $r$ for 5 different cases:
      (i) SFV14, $G/H=2.4$, $Re=1.2\times 10^4$; (ii) SFV14,
      $G/H=3.5$, $Re=1.2\times 10^4$; (iii) SFV20, $G/H=2.4$,
      $Re=1.2\times 10^4$; (iv) SFV20, $G/H=3.5$, $Re=1.2\times 10^4$;
      (v) SFV20, $G/H=3.5$, $R=1.5\times 10^4$.  Different same-colour
      curves correspond to different $y$ positions. (b)
      $S_{2}^{v}/(k^{3/2}r/L)^{2/3}$ versus $r/l_{i}$ (where
      $l_{i}=LR^{-3/4}$ with $R=\sqrt{k}L/\nu$) and
      $S_{2}^{v}/(\langle\varepsilon\rangle r)^{2/3}$ versus $r/\eta$
      (where $\eta = (\nu^{3}/\langle\varepsilon\rangle)^{1/4}$) for
      cases (i) and (ii), i.e. SFV14. (c) Same as (b) but for cases
      (iii), (iv) and (v), i.e. SFV20. (d) Same as (b) and (c) but for
      all cases, SFV14 and SFV20.  }
  \label{fig:S2v GH all compare}
\end{figure}

The theory is based on concepts of inner and outer similarity
which are used to characterise the non-homogeneity of two-point
statistics and on a new inner-outer dissipation equivalence
hypothesis. Future research is required to establish the breadth of
applicability of this hypothesis and of such similarity assumptions
across a variety of non-homogeneous turbulent flows.
Isotropy is not explicitly assumed in the theory, but our analysis of
the data of \cite{chen2021turbulence} suggests that it may be required
at some level. Turbulence production is not explicitely taken into
account either and our data analysis also suggests that the theory
requires modifications when production cannot be neglected. These are
issues which will need to be addressed in future research along with
the question of the particular quantities which may obey similarity at
the inner and/or outer level, and those which may not. Another
important related issue which also requires future research is the
issue of large-scale coherent structures and their effects on or just
relations to similarity and non-homogeneity and turbulence
interscale/interspace transfers. Are they, for example, responsible
for the inner-outer equivalence for turbulence dissipation
as a property of the type of non-homogeneity considered here? How
general is this inner-outer equivalence? We should start thinking in
terms of different classes of non-homogeneity depending, for example,
on the degree of significance of turbulence production and on
differences in coherent structure dynamics in the presence or absence
of turbulence production.  \cite{alves_portela2020} showed that
non-homogeneity and coherent structures are in fact key for the proper
understanding of interscale turbulence transfers in near-field
non-homogeneous turbulent wakes. Future research should probably
combine their approach with the present one and investigate these new
questions.

The data of \cite{chen2021turbulence} which have been used here to
test the theory's new scalings are from three different turbulent
wakes of side-by-side identical square prisms. These three
non-homogeneous turbulent flows differ by their gap ratio $G/H$ and
represent three qualitatively different flow regimes. Good agreement
with the theory's scalings (\ref{eq39:END1}) and (\ref{eq40:END2}) has
been found in all three flows, but far enough from the prisms where
the turbulence is clearly not locally homogeneous but nevertheless
exhibits some indications of local isotropy. Whereas the scalings
(\ref{eq39:END1}) and (\ref{eq40:END2}) collapse the structure
functions rather well in all three flows at these far enough positions
(which are actually not so far as they are only between $10H$ and
$20H$), the intermediate $r^{2/3}$ power law predicted by the theory
is more clearly supported by the data for $S_{2}^{u}$ than for
$S_{2}^{v}$. This may or may not be a Reynolds number effect, which is
another issue needing to be addressed in future studies. However,
there are also clear qualitative differences between $S_{2}^{v}$ for
$G/H=1.25$ and $S_{2}^{v}$ for $G/H=2.4$ and $3.5$ even if
(\ref{eq40:END2}) collapses both cases. These differences point to a
power law for $S_{2}^{v} (r) $ which may be slightly different from
$r^{2/3}$ in the $G/H=1.25$ case. Differences of this sort can be
exploited in future investigations which could lead to a much deeper
understanding of scale-by-scale energy scalings and energy transfers
in non-homogeneous turbulent flows.

Finally it is worth mentioning that (\ref{eq39:END1}) and
(\ref{eq40:END2}) are also able to more or less collapse structure
functions for different cross-stream positions, two far enough
streamwise stations, two flow cases $G/H=2.4$, $3.5$ and two Reynolds
numbers (see figures \ref{fig:S2 GH all compare} and \ref{fig:S2v GH
  all compare}). However, some dependence on inlet conditions remains
and one may need to take into account some dependence of the
similarity functions, particularly $f_v$, on $G/H$.

\vskip 0.5truecm

\noindent
{\bf Acknowledgements} We are grateful to the authors of \citet{chen2021turbulence} for their data, in particular Dr. Christophe Cuvier who has been essential in the setting up of their experiment and data collection.

\vskip 0.5truecm
\noindent
{\bf Funding.} This work was supported by JCV's Chair of Excellence CoPreFlo funded 
by I-SITE-ULNE(grant number R-TALENT-19-001-VASSILICOS); MEL(grant number CONVENTION\_219\_ESR\_06) and Region Hauts de France (grant number 20003862).  

\vskip 0.5truecm
\noindent
{\bf Declaration of interests.} The authors report no confict of interest.

\vskip 1truecm

\appendix
\section{Consistency between the inner scale-by-scale energy
balance and the inner similarity hypothesis}

Using (\ref{eq17:outer-result}) (which we obtained from the outer
scale-by-scale balance in section 3.1) and (\ref{eq21:b1a}), the high
Reynolds number inner scale-by-scale energy balance
(\ref{eq20:SPiRBIG-KHMHdu}) becomes
$$g_{l}^{-1/2}g_{X}\alpha f_{iX}(r/l_{i}) +g_{l}^{-1/2}g_{3}\gamma
f_{i3}(r/l_{i}) +g_{l}^{-1/2}g_{p}\beta f_{ip}(r/l_{i}) =
$$
\begin{equation}
-1 + C_{\varepsilon}^{-1}{\bf \nabla}^{2}_{r/l_{i}} f_{i2}(r/l_{i})
\label{eq22:SPiRBIG-KHMHdu-better}
\end{equation}
in terms of the dimensionless proportionality constants $\alpha =
(V_{oX}/V_{o2})^{3}/C_{\varepsilon}$, $\beta =
(V_{op}/V_{o2})^{3}/C_{\varepsilon}$ and $\gamma =
(V_{o3}/V_{o2})^{3}/C_{\varepsilon}$ (some of these constants, though
not all, could in principle be zero). Irrespective of which
transport/transfer term we keep in this high Reynolds number inner
energy budget, i.e. which of $g_{l}^{-1/2}g_{X}$, $g_{l}^{-1/2}g_{3}$
and $g_{l}^{-1/2}g_{p}$ we let tend to zero in the limit $R\to
\infty$, we end up with a function of $r/l_{i}$ being equal to $-1 +
C_{\varepsilon}^{-1}{\bf \nabla}^{2}_{r/l_{i}} f_{i2}(r/l_{i})$ which
is possible either if $C_{\varepsilon}$ is constant, i.e. independent
of ${\bf X}$, or if ${\bf \nabla}^{2}_{r/l_{i}} f_{i2}(r/l_{i})=0$
which means that the longitudinal second order structure function
should be a harmonic function of $r$, which is not realistic. As we
want a theory for non-homogeneous turbulence where $C_{\varepsilon}$
depends on ${\bf X}$, we need to modify some of the inner similarity
assumptions (\ref{eq11:SP3i}), (\ref{eq12:SPXi}) and/or
(\ref{eq13:SPpi}).

Firstly, we keep the interscale turbulence energy transfer in the high
Reynolds number inner scale-by-scale energy balance. This requires
 \begin{equation}
g_{l}^{-1/2} (R)g_{3} (R) = Const 
\label{eq23:cb2}
 \end{equation}
 independent of $R$. Secondly, we take $g_{l}^{-1/2}g_{X} = Const$ and
 $g_{l}^{-1/2} g_{p} = Const$ to allow for the possibility of both or
 either of the turbulent transport and the velocity-pressure gradient
 correlation terms to be present in the high Reynolds number inner
 scale-by-scale energy balance. (These terms can be made to be absent
 from this balance by artificially setting $\alpha$ and/or $\beta$
 equal to $0$ respectively.)  We chose to modify the inner similarity
 forms of these two terms as they concern statistics which do not only
 involve two-point velocity differences whereas this is not the case
 of the interscale transfer rate term which does. Looking at
 (\ref{eq22:SPiRBIG-KHMHdu-better}), the modification that we are
 forced to make must ensure that $C_{\varepsilon} (1 + \alpha f_{iX}
 +\gamma f_{i3} + \beta f_{ip})$ is independent of ${\bf X}$
 (explicitely). We must therefore replace $f_{iX}$ and $f_{ip}$ in
 (\ref{eq12:SPXi}) and (\ref{eq13:SPpi}) by the following functions of
 $r/l_{i}$ {\it and} ${\bf X}$:
\begin{equation}
f_{iX} (r/l_{i}, {\bf X}) = {F_{iX}(r/l_{i})\over C_{\varepsilon}({\bf
    X})} - A - B f_{i3} (r/l_{i})
\label{eq24:SPXnew}
\end{equation}
\begin{equation}
f_{ip} (r/l_{i}, {\bf X}) = {F_{ip}(r/l_{i})\over C_{\varepsilon}({\bf
    X})} - C - D f_{i3} (r/l_{i})
\label{eq25:SPpnew}
\end{equation}
where $A$, $B$, $C$ and $D$ are dimensionless constants and $F_{iX}$
and $F_{ip}$ are functions of $r/l_{i}$ only. Note that if the
turbulent transport term is not present in the high Reynolds number
inner scale-by-scale energy balance then $F_{iX}=0$ and $A=B=0$, and
if the velocity-pressure gradient correlation term is not present in
that balance then $F_{ip}=0$ and $C=D=0$. All possibilities are
therefore covered.

With (\ref{eq24:SPXnew}) and (\ref{eq25:SPpnew}), the requirement that
$C_{\varepsilon} (1 + \alpha f_{iX} +\gamma f_{i3} + \beta f_{ip})$
must be independent of ${\bf X}$, and in fact equal to ${\bf
  \nabla}^{2}_{r/l_{i}} f_{i2}(r/l_{i})$, yields $\gamma = \alpha B +
\beta D$, $1=\alpha A + \beta C$ and $\alpha F_{iX} + \beta F_{ip} =
      {\bf \nabla}^{2}_{r/l_{i}} f_{i2}(r/l_{i})$. Taking into account
      (\ref{eq11:SP3i}), (\ref{eq12:SPXi}), (\ref{eq13:SPpi}) and
      (\ref{eq14:dissipation}), this latter equation represents the
      following high Reynolds number balance for $r\ll l_{o}$:
$$
{\bf \nabla}_X \cdot \langle {\bf u}_X^{\prime} (\delta u_{1})^{2}\rangle + {\bf
  \nabla}_r \cdot \langle \delta {\bf u} (\delta u_{1})^{2}\rangle
+2\langle \delta u_{1} {\partial \over \partial X_{1}} \delta p\rangle=
$$
\begin{equation}
-\varepsilon_{1} + {V_{o2}^{3}\over l_{o}}{\bf \nabla}^{2}_{r/l_{i}}
f_{i2}(r/l_{i})
\label{eq26:epsiloninnerbalance}
\end{equation}

Not only does the theory not lead to unrealistic conclusions with the
modifications (\ref {eq24:SPXnew}) and (\ref{eq25:SPpnew}) of the
inner similarity forms, it also leads to this, arguably interesting,
high Reynolds number balance (\ref{eq26:epsiloninnerbalance}). The
possibility that this balance may in fact involve only ${\bf \nabla}_X
\cdot \langle {\bf u}_X^{\prime} (\delta u_{1})^{2}\rangle$ and ${\bf
  \nabla}_r \cdot \langle \delta {\bf u} (\delta u_1)^2\rangle$ or
only ${\bf \nabla}_r \cdot \langle \delta {\bf u} (\delta
u_1)^2\rangle$ and $2\langle \delta u_{1} {\partial \over \partial
  X_{1}} \delta p\rangle$ exists and is covered by our approach (just
take $F_{ip}=0$ and $C=D=0$ in the former case and $F_{iX}=0$ and
$A=B=0$ in the latter). One should not interpret
(\ref{eq26:epsiloninnerbalance}) to necessarily mean that interscale
energy transfer and turbulence dissipation are balanced by both
turbulent transport in space and the velocity-pressure gradient
correlation term. They are balanced by at least one or the other or
both. This is a consequence of the choice we made not to modify the
inner similarity form of the interscale transfer rate and keep
$f_{i3}$ as function of $r/l_{i}$ only.


\bibliographystyle{jfm}
\bibliography{jfmRefs.bib}

\end{document}